\documentclass[11pt]{article}

\usepackage[utf8]{inputenc}
\usepackage[T1]{fontenc}
\usepackage{fixltx2e}
\usepackage{graphicx}
\usepackage{longtable}
\usepackage{float}
\usepackage{wrapfig}
\usepackage{soul}
\usepackage{textcomp}
\usepackage{marvosym}
\usepackage[nointegrals]{wasysym}
\usepackage{latexsym}
\usepackage{amssymb}
\usepackage{hyperref}
\usepackage{numprint}
\usepackage[height=24cm]{geometry}
\usepackage[english]{babel}
\usepackage{amsmath,amsthm}
\usepackage{algorithmicx}
\usepackage{algorithm}
\usepackage{algpseudocode}
\usepackage{mathrsfs}
\usepackage{comment}
\usepackage{subfig}
\usepackage{xcolor}
\usepackage{authblk}
\usepackage{wrapfig}
\graphicspath{{FIGURES/}}
\newcommand{\R}{\mathbb{R}}
\newcommand{\E}{\mathbb{E}}
\newcommand{\Te}{\mathcal{T}}

\newcommand{\LL}{\mathbf{L}}
\newtheorem{definition}{Definition}
\newtheorem{example}{Example}

\providecommand{\keywords}[1]
{
  \small	
  \textbf{{Keywords---}} #1
}

\title{Gibbsian T-tessellation model for agricultural landscape characterization}
\author[1]{Katarzyna Adamczyk-Chauvat}
\author[2]{Mouna Kassa}
\author[1]{Kiên Kiêu}
\author[3]{Julien Papaïx}
\author[4]{Radu S. Stoica}

\date{}
\hypersetup{
  pdfkeywords={},
  pdfsubject={},
  pdfcreator={Emacs Org-mode version 7.9.3f}}

\affil[1]{MaIAGE, INRAE, Université Paris-Saclay, Jouy-en-Josas, France}
\affil[2]{INSA, Rennes, France}
\affil[3]{BioSP, INRAE, Avignon, France}
\affil[4]{IECL, Université de Lorraine, Nancy, France}
\begin{document}

\maketitle
\begin{abstract}
A new class of planar tessellations, referred to as T-tessellations, was introduced in (\cite{ref/10}).
A completely random T-tessellation model (CRTT)  was proposed, and its Gibbsian variants were discussed. A general 
simulation algorithm of Metropolis-Hastings-Green type was derived for model simulation, involving three local transformations 
of T-tessellations.

The current paper focuses on statistical inference for Gibbs models of T-tessellations.
Statistical methods derived from point pattern analysis are  implemented
on the example of three agricultural landscapes approximated by T-tessellations. 
The choice of model statistics is guided by their capacity to highlight the differences
between the landscape patterns. Model parameters are estimated by the Monte Carlo Maximum
Likelihood method, yielding a baseline for landscapes comparison. In the last part of the paper, 
a global envelope test based on the empty-space function is proposed for assessing the goodness-of-fit 
of the model.

\end{abstract}

\setcounter{tocdepth}{3}

\section{Introduction}

\begin{figure}
\centering
{\includegraphics[width=.35\textwidth]{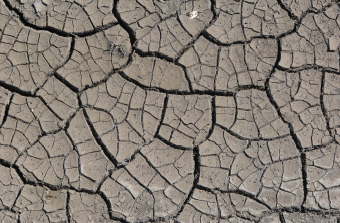}}\quad\quad
{\includegraphics[width=.35\textwidth]{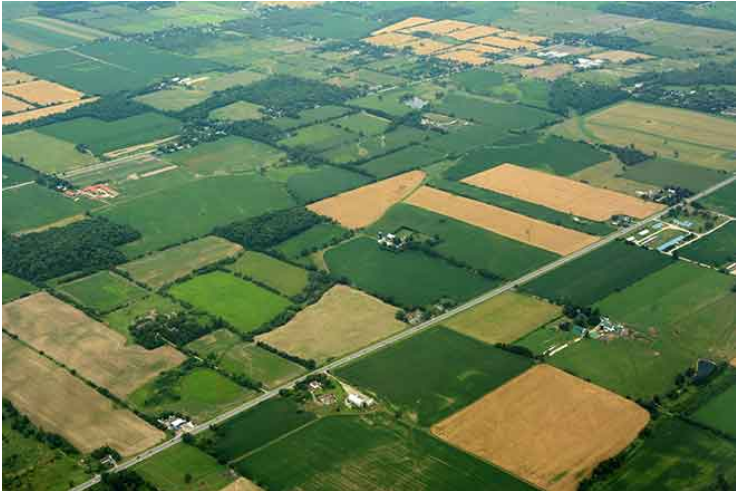}}

\vspace*{.2cm}
{\includegraphics[width=.35\textwidth]{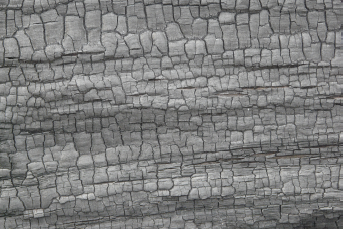}}\quad\quad
{\includegraphics[width=.35\textwidth]{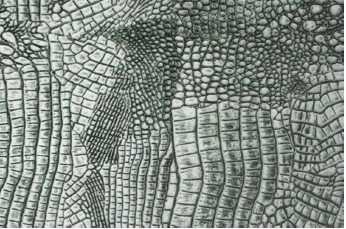}}
\caption{Examples of spatial patterns that may be approximated by a T-tessellation. From upper left panel to lower right one: a cracked soil, an agricultural landscape in Colorado, a texture of burnt wood, a fragment of alligator skin.}
\label{ExTes}
\end{figure}

Tessellations are mathematical representations of space division into cells. This paper focuses
on a class of tessellations with internal vertices with three incident edges,
two of them subtending a straight angle. Such tessellations are referred to as T-tessellations and
may represent various spatial patterns, like those illustrated in Figure (\ref{ExTes}): cracked soil, agricultural landscapes, burnt  wood, 
patterns on a reptile's skin. Random T-tessellation models allow for spatial variability of such patterns, generating  cells
with different sizes and shapes. One of the first  random T-tessellation models was  proposed
by Gilbert (\cite{ref/24}) for representing needle-shaped crystals. A variant of the model providing rectangular
cells was studied by Miles and Mackisack (\cite{ref/13}). The stable with respect to iteration (STIT) tessellation model was proposed 
by Nagel and Weiss  (\cite{ref/26}) for modeling random crack networks.  Another example of a random T-tessellation was proposed by
Arak, Clifford and Surgailis (\cite{ref/25}) in their paper on a random graph model. For the particular
choice of the parameters, the model yields T-tessellation patterns.

More recently, a Completely Random T-Tessellation (CRTT)  model was introduced by Kiêu et al. (\cite{ref/10}).
The  model, based on Poisson line process, was considered as a reference measure on a set of T-tessellations.
Gibbs T-tessellations models can be constructed by means of a probability density w.r.t. the CRTT reference measure.
 The density involves a set of tessellation statistics the model should account for. The tessellations
with large (resp. small) values of the targeted statistics are favored by an appropriate tuning
of model parameters. The possibility of the model to control various tessellation features makes it
attractive for representing a broad scope of the observed T-tessellation patterns.
A Metropolis-Hasting-Green  algorithm for model simulation was derived in (\cite{ref/10}) and implemented in the C++ library (\cite{ref/1}).

This paper focuses on statistical inference for Gibbsian T-tessellation models. The models parameters were
estimated by Monte Carlo Maximum Likelihood (MCML). The estimation method was tested on the example of three agricultural landscapes,
approximated by T-tessellations. The models fitted to the landscapes aimed at finding the features responsible for
the observed contrasts of landscape structures. The goodness-of-fit of the models was assessed by a global envelope test based on the
empty-space  function $F$. 

The paper is structured as follows. Section (2) describes landscape datasets and  presents the algorithm of landscape approximation
by a T-tessellation. Section (3) defines the Gibbsian T-tessellation model. The principle of MCML is outlined and the properties
of the estimates are investigated by  simulations. Section (4) presents the results of  model fit to the three datasets
and points out the main differences between the landscapes. The goodness-of-fit test is carried out for each dataset. Finally, 
Section (5)  contains the concluding remarks and highlights further perspectives.

\section{From landscape data to a T-tessellation}

Agricultural landscapes exhibit patterns that can be reasonably approximated by T-tessellations. Indeed,
as illustrated in Figure (\ref{fig:5}), most of the vertices of polygons representing agricultural fields are of the T-type. 
Although the T-vertices are dominant, actual agricultural landscapes are not T-tessellations and must be modified
to match the definition of T-tessellation. In the following, an algorithm for transforming a landscape into a T-tessellation
is proposed and tested on three datasets. 

\subsection{Datasets}

\begin{figure}
\centering
\subfloat[]{\includegraphics[width=.31\textwidth]{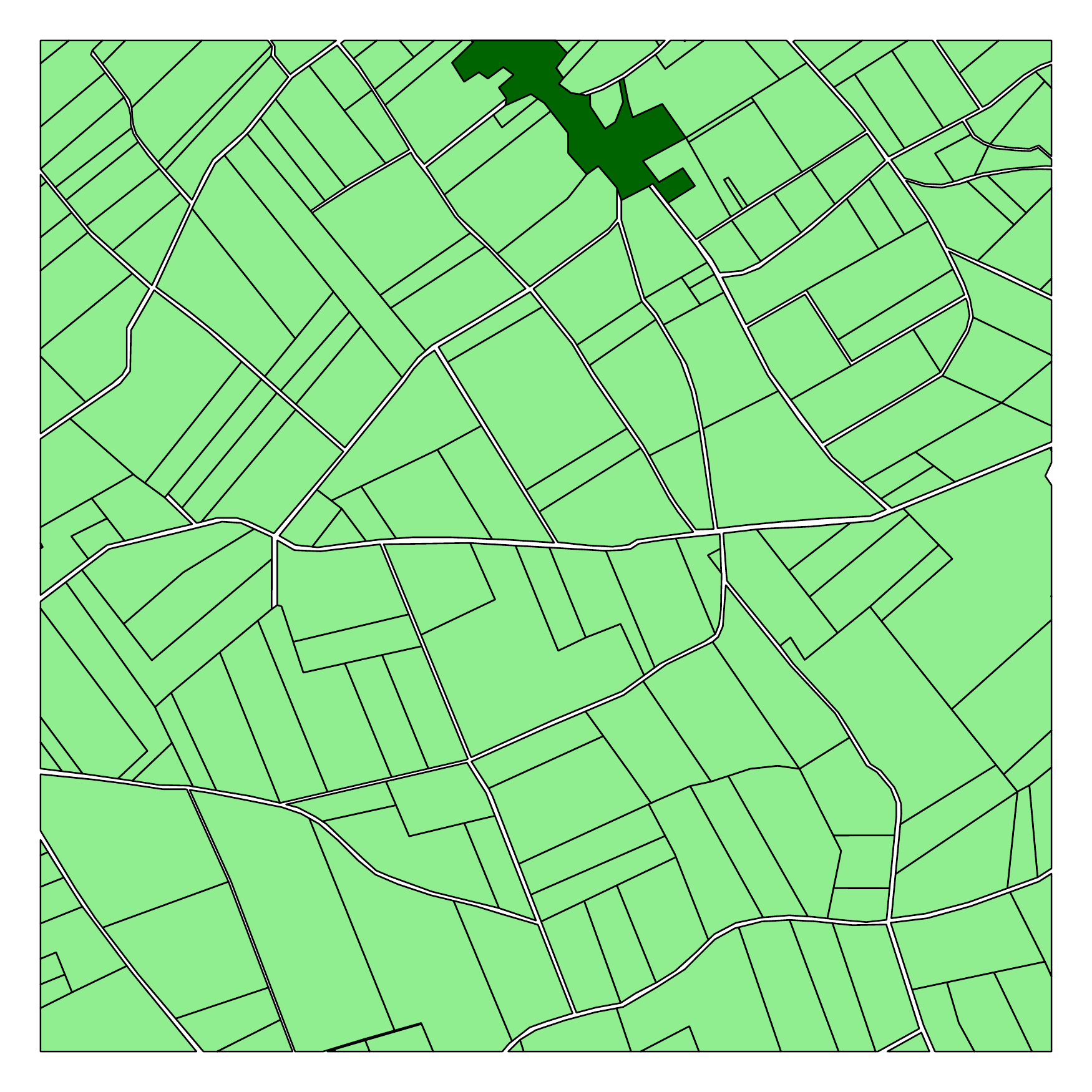}}\quad
\subfloat[]{\includegraphics[width=.31\textwidth]{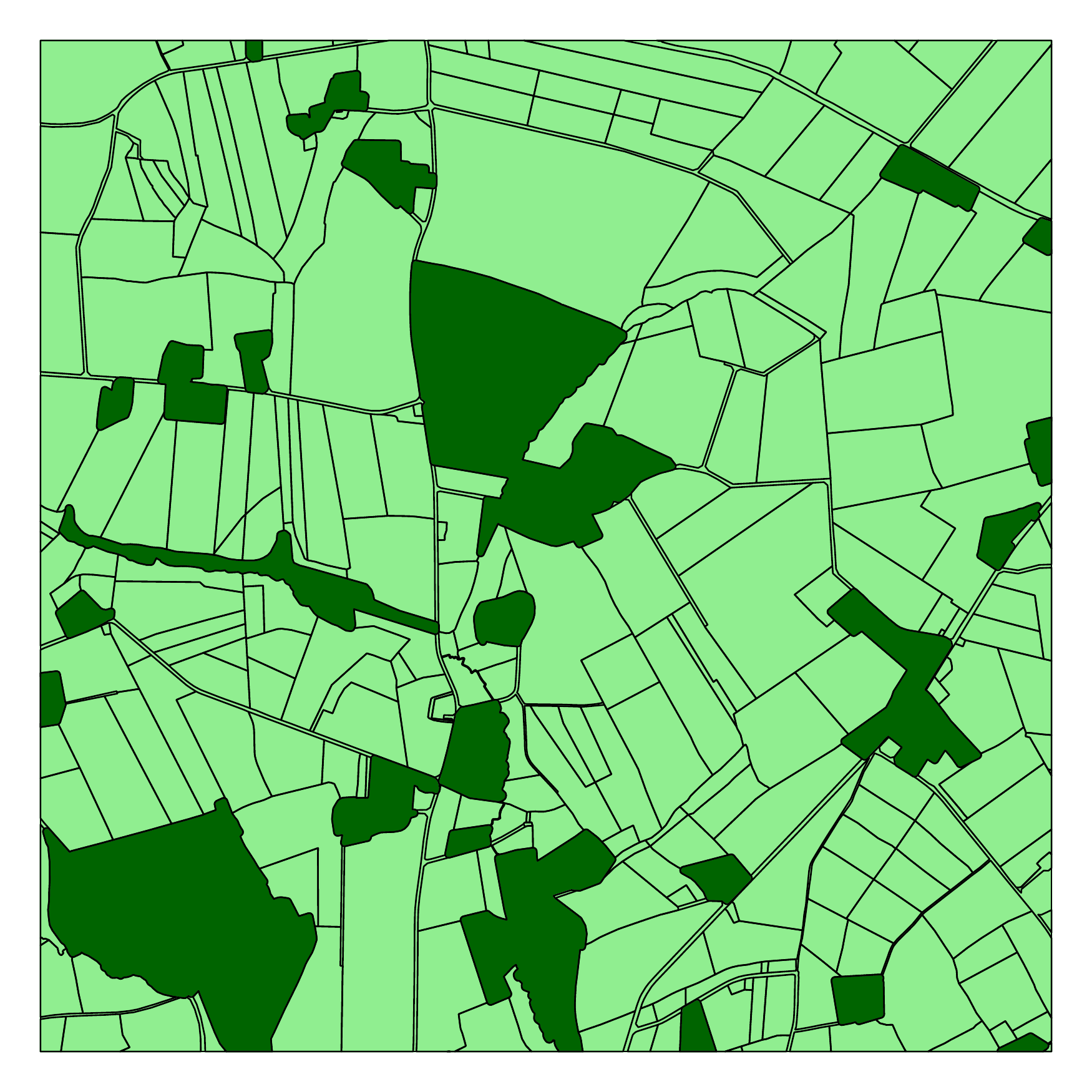}}\quad
\subfloat[]{\includegraphics[width=.31\textwidth]{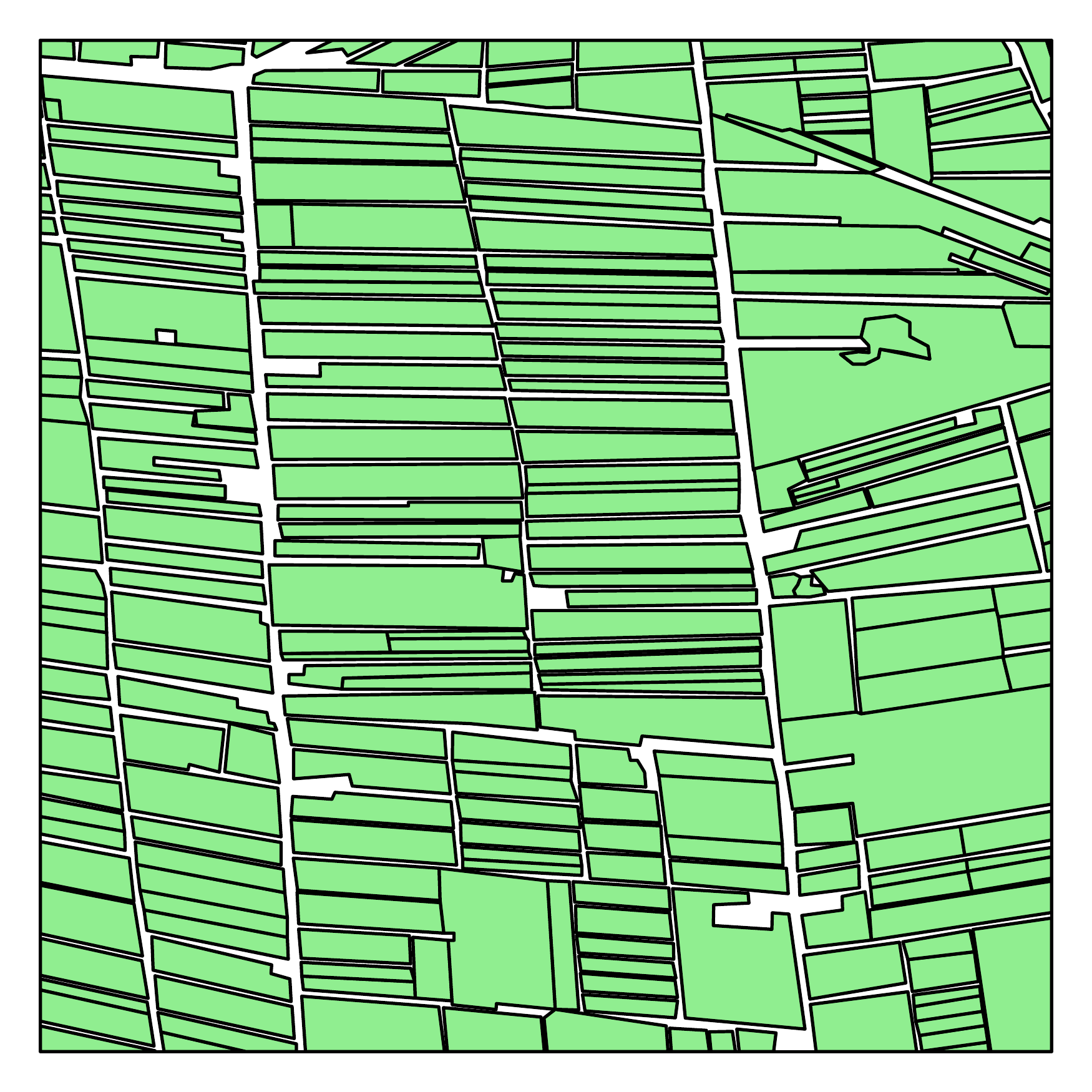}}\quad
\subfloat[]{\includegraphics[width=.31\textwidth]{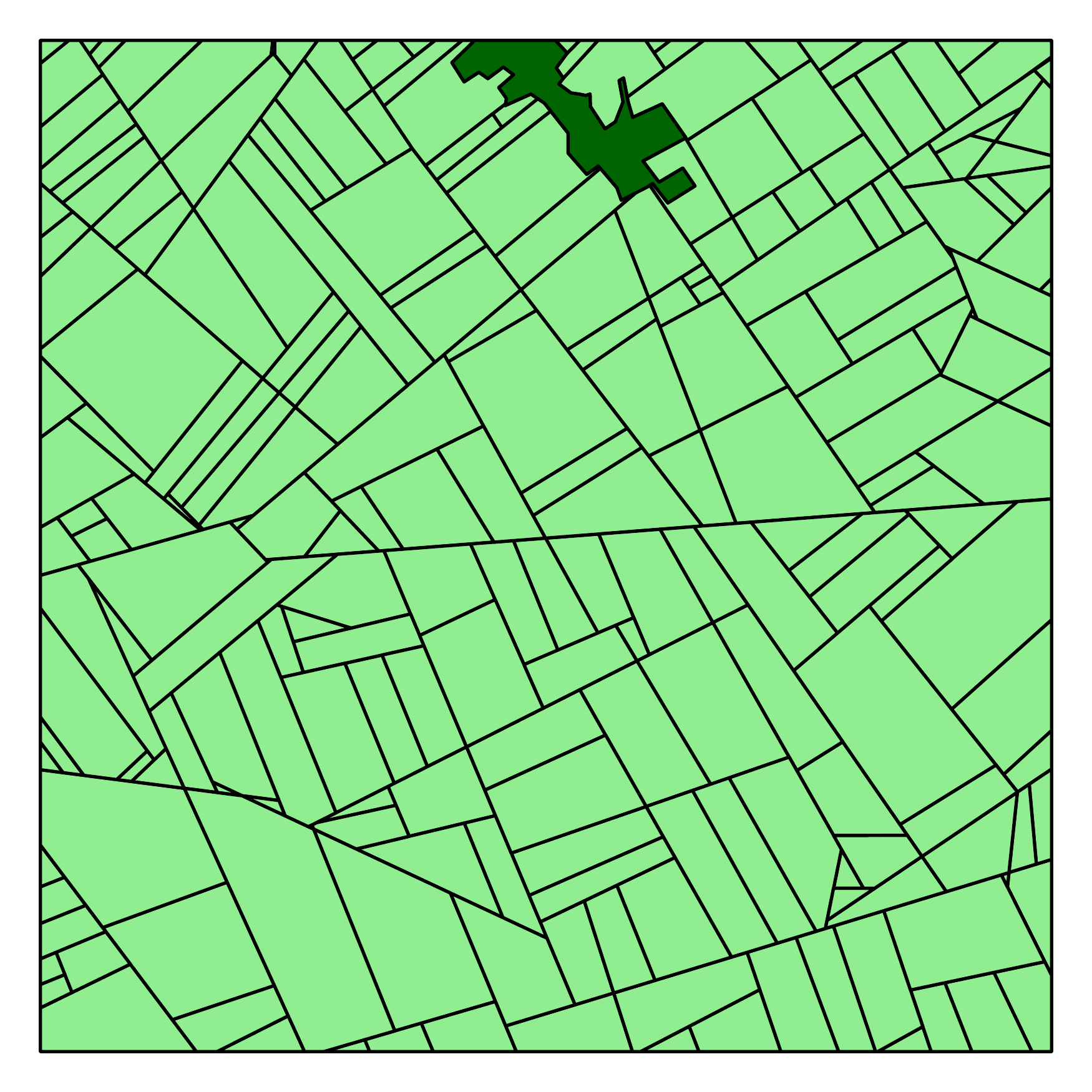}}\quad
\subfloat[]{\includegraphics[width=.31\textwidth]{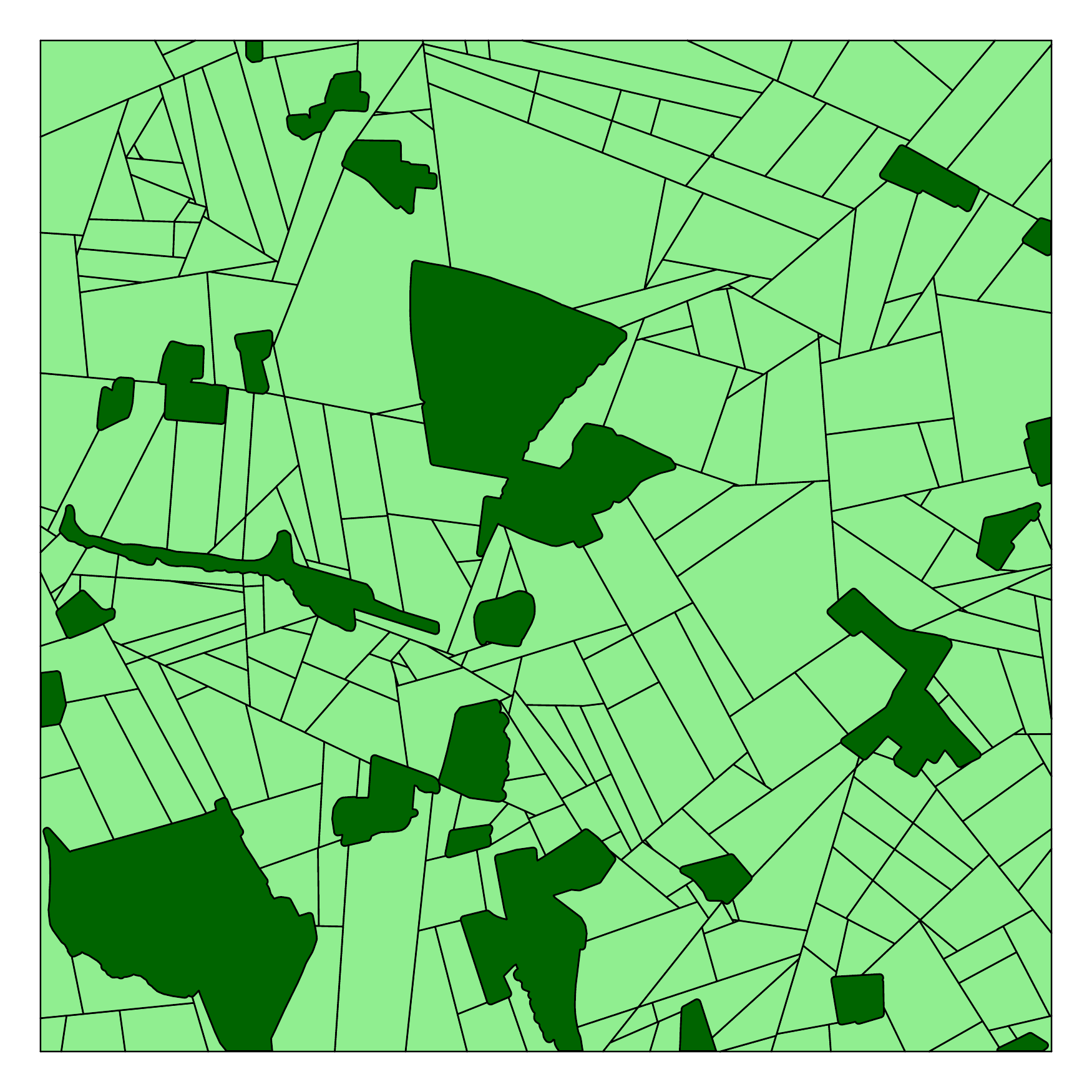}}\quad
\subfloat[]{\includegraphics[width=.31\textwidth]{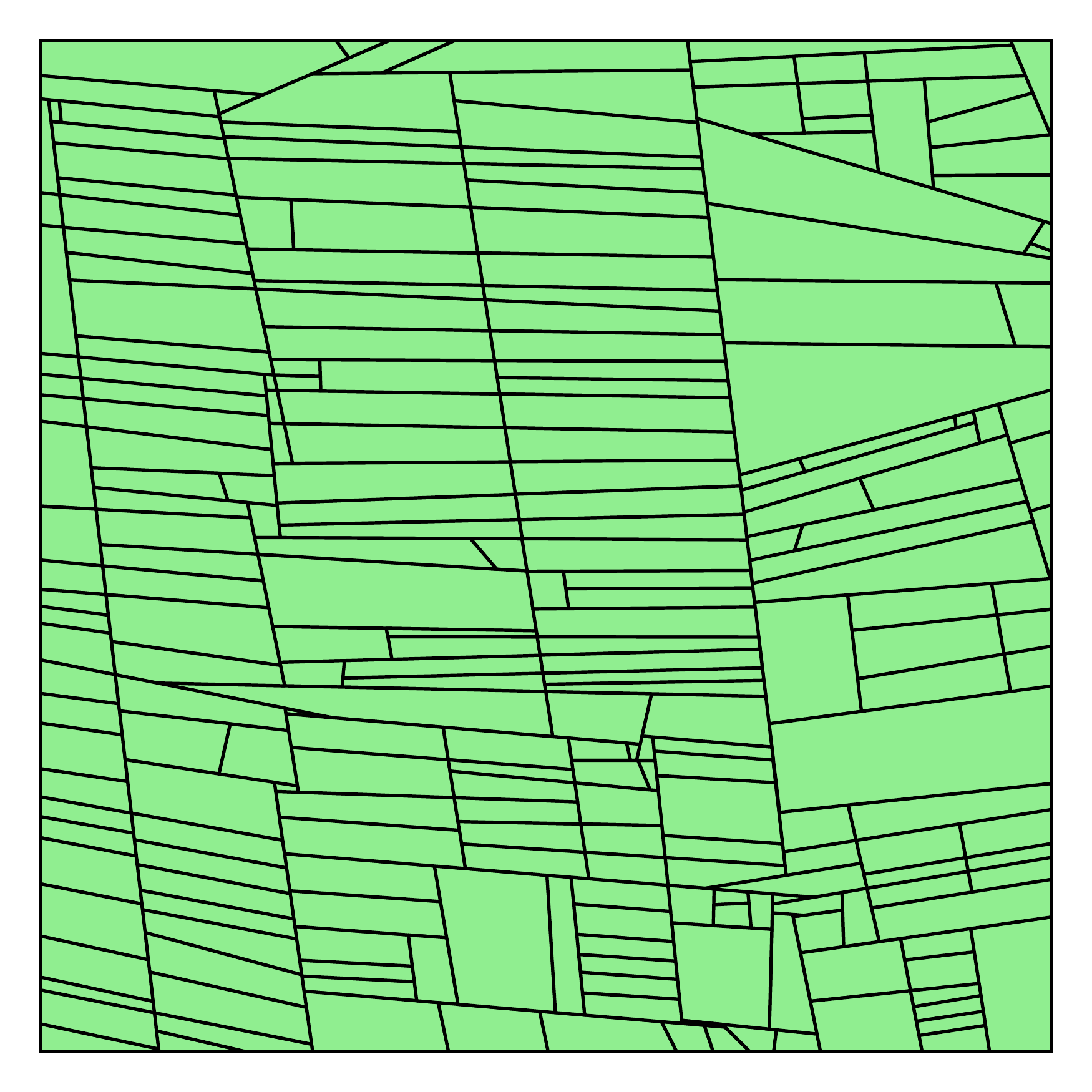}}\quad

\caption{Upper row: landscapes of Selommes (a); Kervidy (b); and BVD (c).  Light green polygons depict agricultural plots, 
dark green polygons stand for non-agricultural areas (mainly woods and villages). Lower row:
landscapes approximated by T-tessellations.}
\label{fig:5}
\end{figure}

Three French agricultural landscapes were selected for the study:
the landscape of Selommes (Centre-Val de Loire), an area of intensive cereal crop cultivation;
the landscape of Kervidy (Brittany), a region characterized by mixed crop-livestock farming systems; and the landscape 
of the Basse Valée de la Durance (BVD) in Provence-Alpes-Côte d'Azur region, specialized in apple orchards. Different types of agricultural production characterizing the three regions are in part responsible for the  variability of landscape patterns observed in Figure (\ref{fig:5}). A more
detailed presentation of the datasets and the related agroecological research problems can be found
in (\cite{ref/20}) for the Selommes landscape, in (\cite{ref/12}) for the Kervidy landscape and in (\cite{ref/29}) for the BVD landscape.
The landscapes were restricted to the domains encompassing a comparable number of fields: $230$ for Selommes, 
$245$ for BVD and $287$ for Kervidy.

\subsection{Landscape approximation algorithm}
\begin{figure}
  \centering
  \includegraphics[width=.45\textwidth]{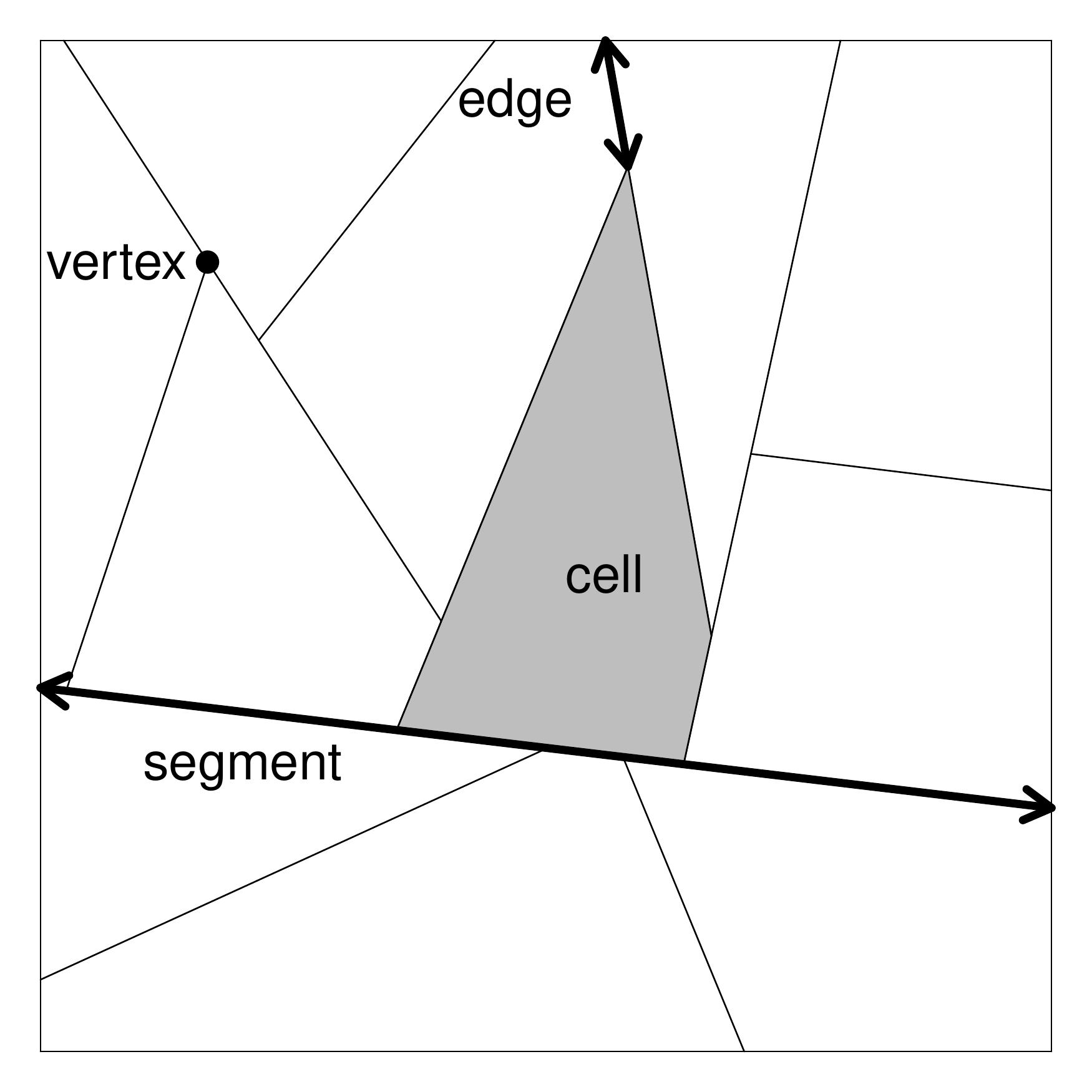}
  \caption{T-tessellation of a squared window.}
  \label{fig:T:tessellation}
\end{figure}

A polygonal tessellation of a squared domain $W$ is illustrated in Figure  (\ref{fig:T:tessellation}).
The main components of the tessellation are highlighted: the cells, the edges (non-empty
intersections of two cells), the vertices (non-empty intersections of three cells) and the
segments (maximal unions of the aligned and contiguous edges). 

\begin{definition}
A polygonal tessellation of a  bounded domain $W\subset\R^2$ is called a T-tessellation if the following conditions
are simultaneously fulfilled:
\begin{description}
\item[Condition 1] Each internal vertex of a tessellation has exactly three adjacent edges, two of them subtending the angle of $\pi$.
\item[Condition 2] There is no pair of distinct and aligned segments.
\end{description}
\label{def}
\end{definition}

The purpose of the approximation algorithm is to construct a T-tessellation that closely matches the input set of polygons.
The algorithm requires the coordinates of polygons  and the coordinates of  the tessellated domain, composed from an outer polygon and polygons-holes.
The consecutive steps of the algorithm are described below.

\begin{figure}
\centering
\subfloat[]{\includegraphics[width=.27\textwidth]{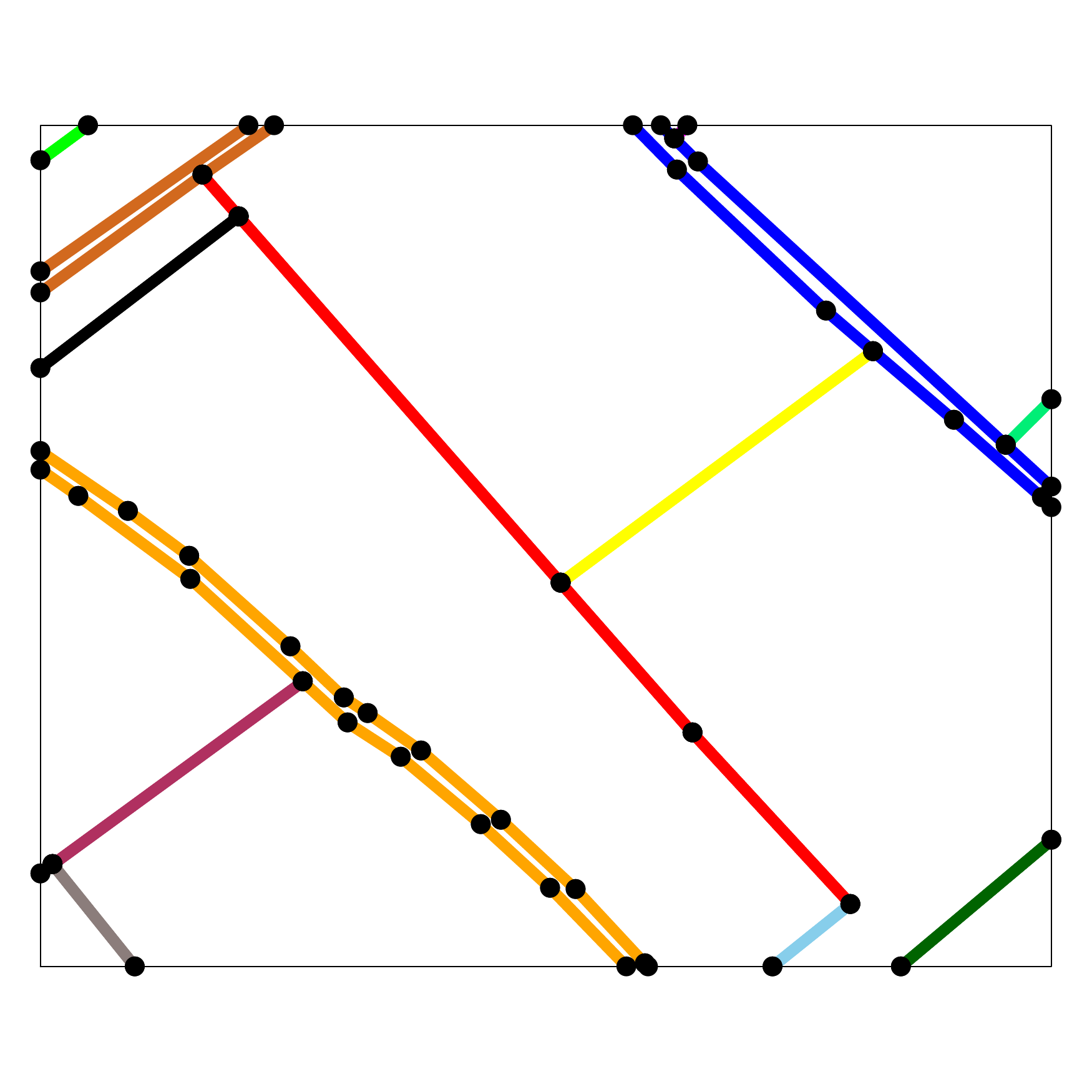}}\quad
\subfloat[]{\includegraphics[width=.27\textwidth]{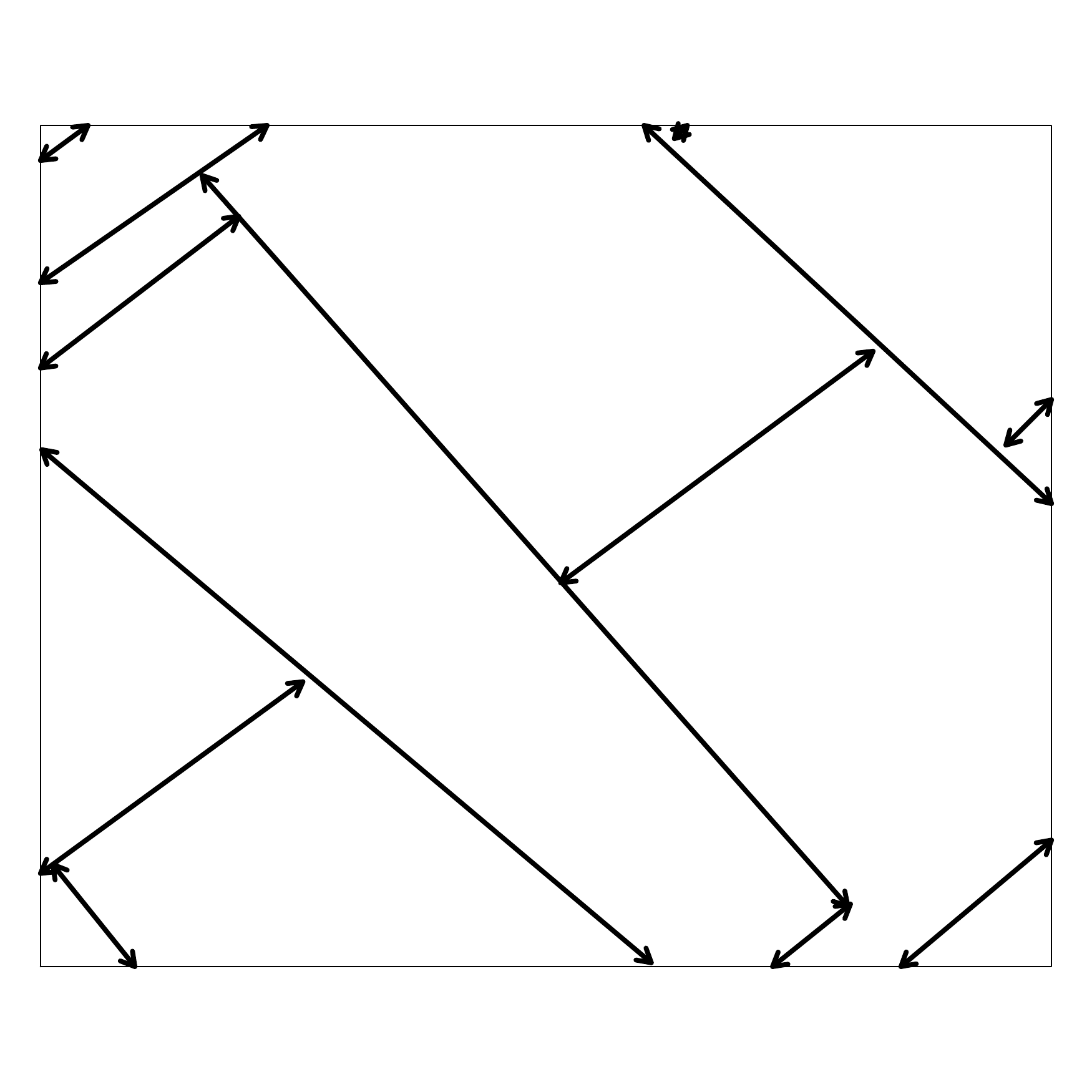}}\quad
\subfloat[]{\includegraphics[width=.27\textwidth]{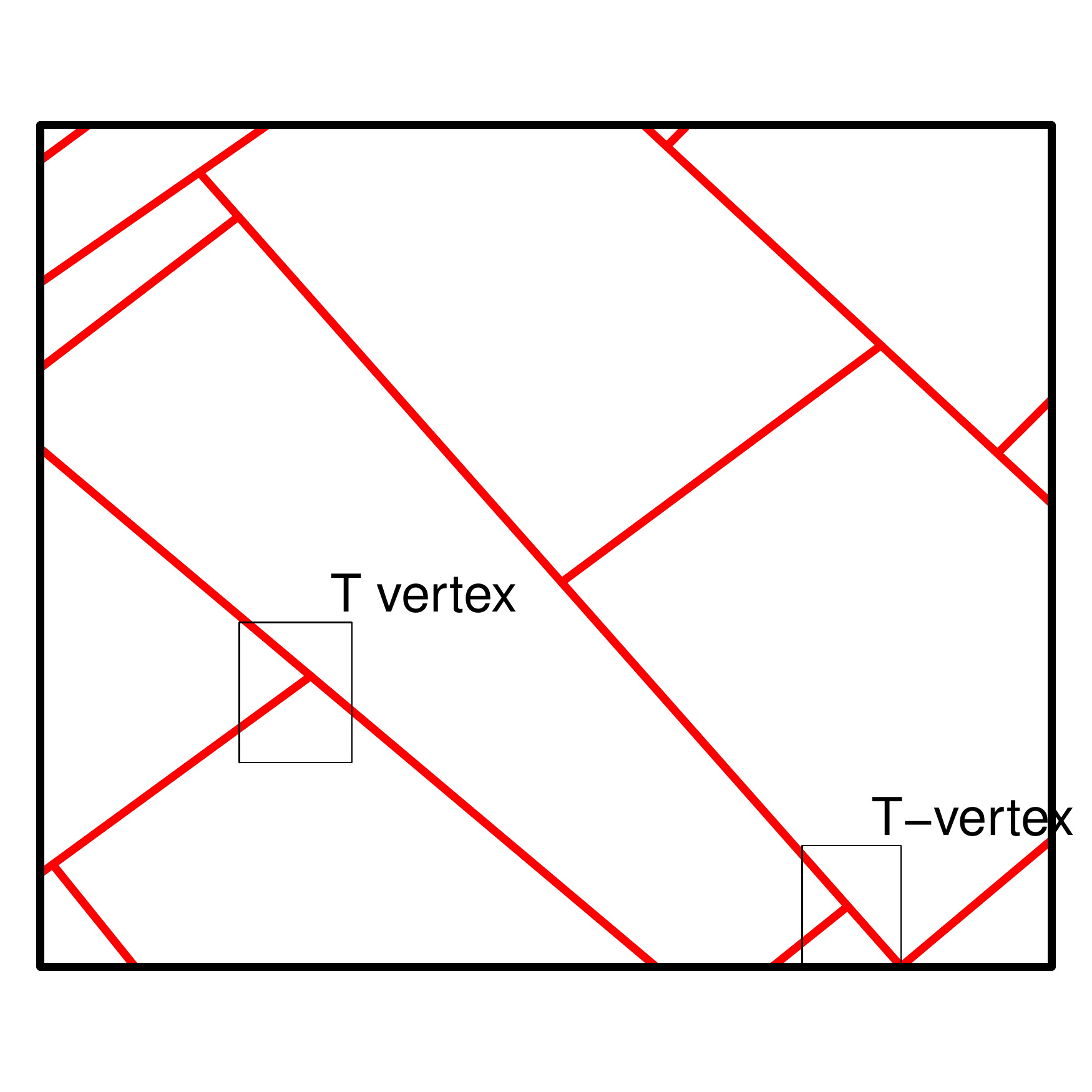}}
\caption{A fragment of the Selommes landscape transformed into a T-tessellation: (a) clusters of closed and aligned polygons sides; (b) tessellation spanned by representative segments; (c) tessellation after removal of the I- and L-vertices.}
\label{approxAlgo}
\end{figure}

\paragraph{Grouping polygons sides (Fig. (\ref{approxAlgo}a))} 

The first step of the algorithm consists of grouping the sides of polygons  by  hierarchical clustering. Two sides  belong to the same group if they are spatially close and if their orientation is similar. The dissimilarity measure that takes  these two aspects into account is defined as the sum of two terms: the
area of the convex envelope of two sides, denoted by $A(C(.))$, and  the smallest Euclidean distance between
the sides, denoted by $d_{\min}(.)$:
\begin{equation}
\label{eq:side:dissimilarity}
  d(c_i,c_j)= \frac{A(C(c_i,c_j))}{l(c_i)l(c_j)}+\frac{2 d_{\min}(c_i,c_j)}{{l(c_i)+l(c_j)}},
\end{equation}
The first term becomes large when the sides $c_i$ and $c_j$  are far apart and/or perpendicularly oriented; the second one accounts for the distant sides belonging to the same straight line.
To make the two components of the criterion independent of the metric, they are normalized  by the product of the lengths of the sides and 
by their average length, respectively. The classification of the sides is based on the single linkage method.
In this method, the distance between two groups of sides is defined as the minimal dissimilarity between the pairs of sides from each group.
At each step of the algorithm, the two closest groups are merged, starting from the groups containing single sides.

\paragraph{Choice of representative segments (Fig. (\ref{approxAlgo}b))} Each group of sides is then replaced by a representative segment: a regression line is fitted to the endpoints of the grouped sides. The projection of each endpoint on the regression line is calculated. A representative segment is the smallest segment on the regression line, including the projections of sides endpoints.

\paragraph{Transformation into T-tessellation (Fig. (\ref{approxAlgo}c))} 

The representative segments are inserted into a tessellation domain, defining a tessellation that is not yet a T-tessellation. 
It contains the I-vertices with one incident edge, L-vertices with two incident edges, and X-vertices
formed by the crossing segments. To remove an I-vertex, the edge can be removed or lengthened until it meets another edge. In order to choose
between these two modifications, a variation of total internal length  is calculated: the modification that induces the smallest length variation is selected.
The same rule also governs the order of removing the I-vertices: at each iteration, the I-vertex whose suppression induces the smallest variation in length
is selected. The approach for removing the L-vertices is quite similar: one of two incident edges is lengthened until it meets another edge.
The edge that induces the smallest increase in total internal length is selected. The approach for fixing the order of removal of L-vertices 
is the same as for I-vertices.

In order to remove an X-vertex from a tessellation, one of two segments passing through the vertex is broken
into two halves  at the vertex location. The end of a segment half lying at the vertex location is moved along the other incident segment
by a small distance. The procedure is repeated until there is no X-vertex left in the tessellation.

\begin{figure}[b]
\centering
\subfloat[]{\includegraphics[width=.27\textwidth]{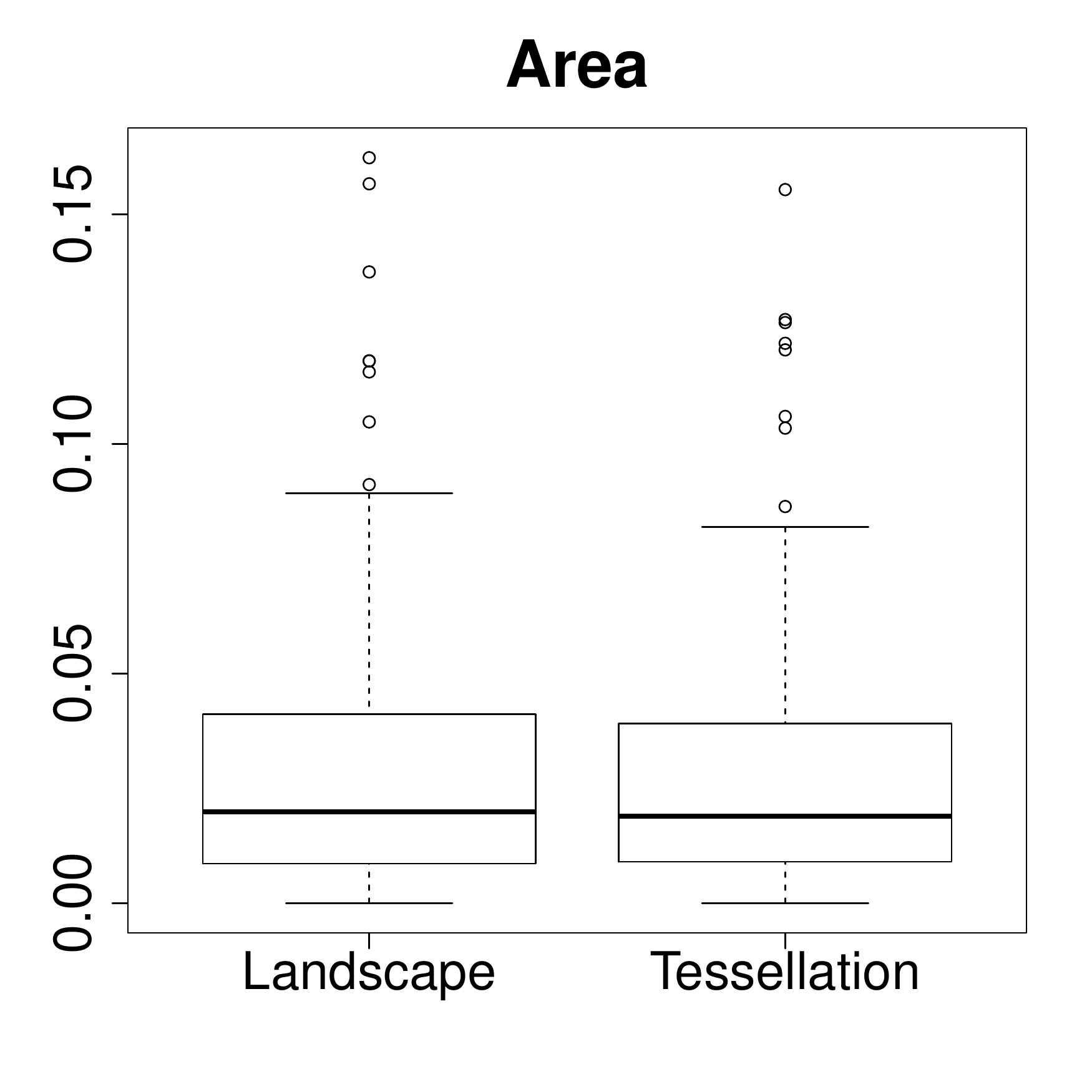}}\quad
\subfloat[]{\includegraphics[width=.27\textwidth]{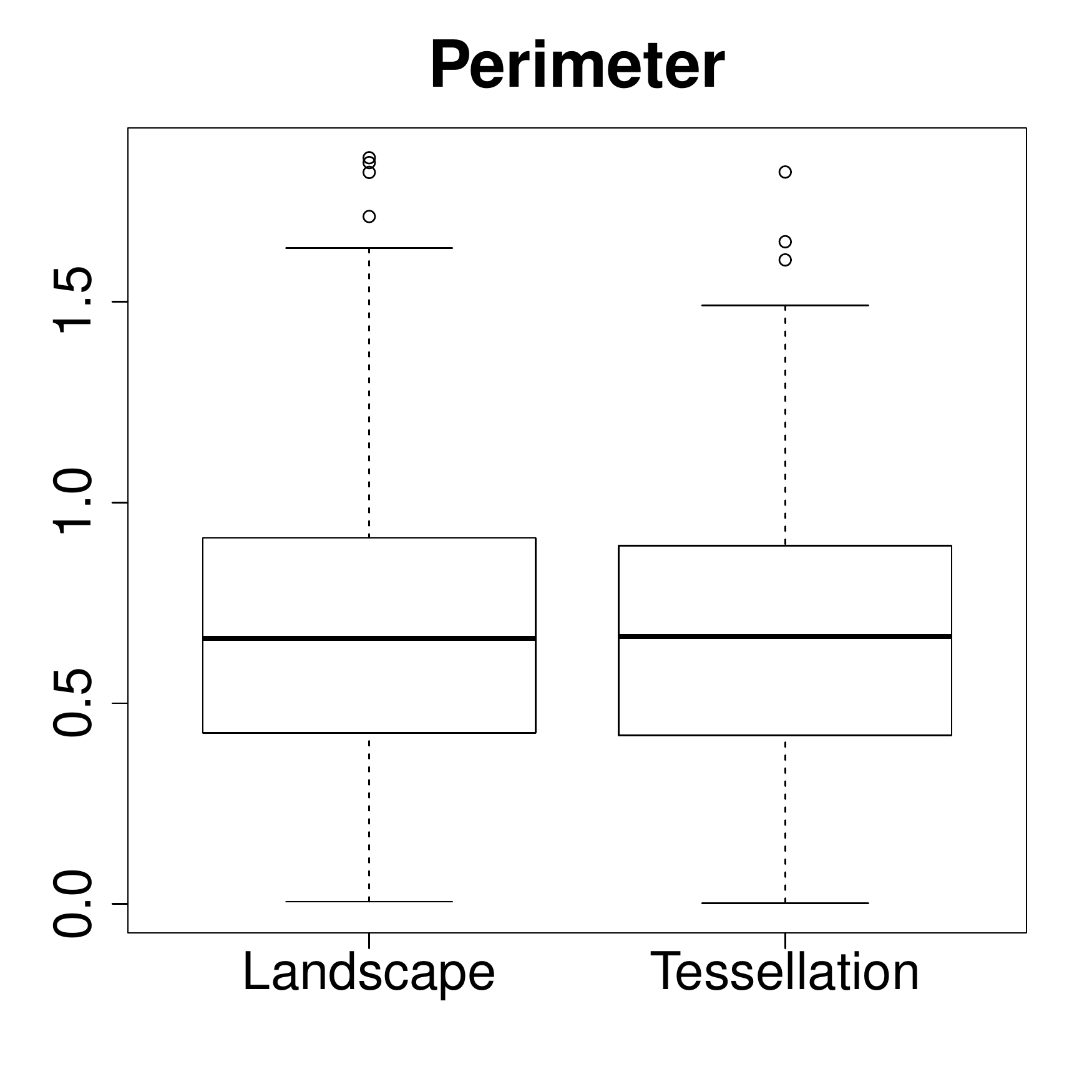}}\quad
\subfloat[]{\includegraphics[width=.27\textwidth]{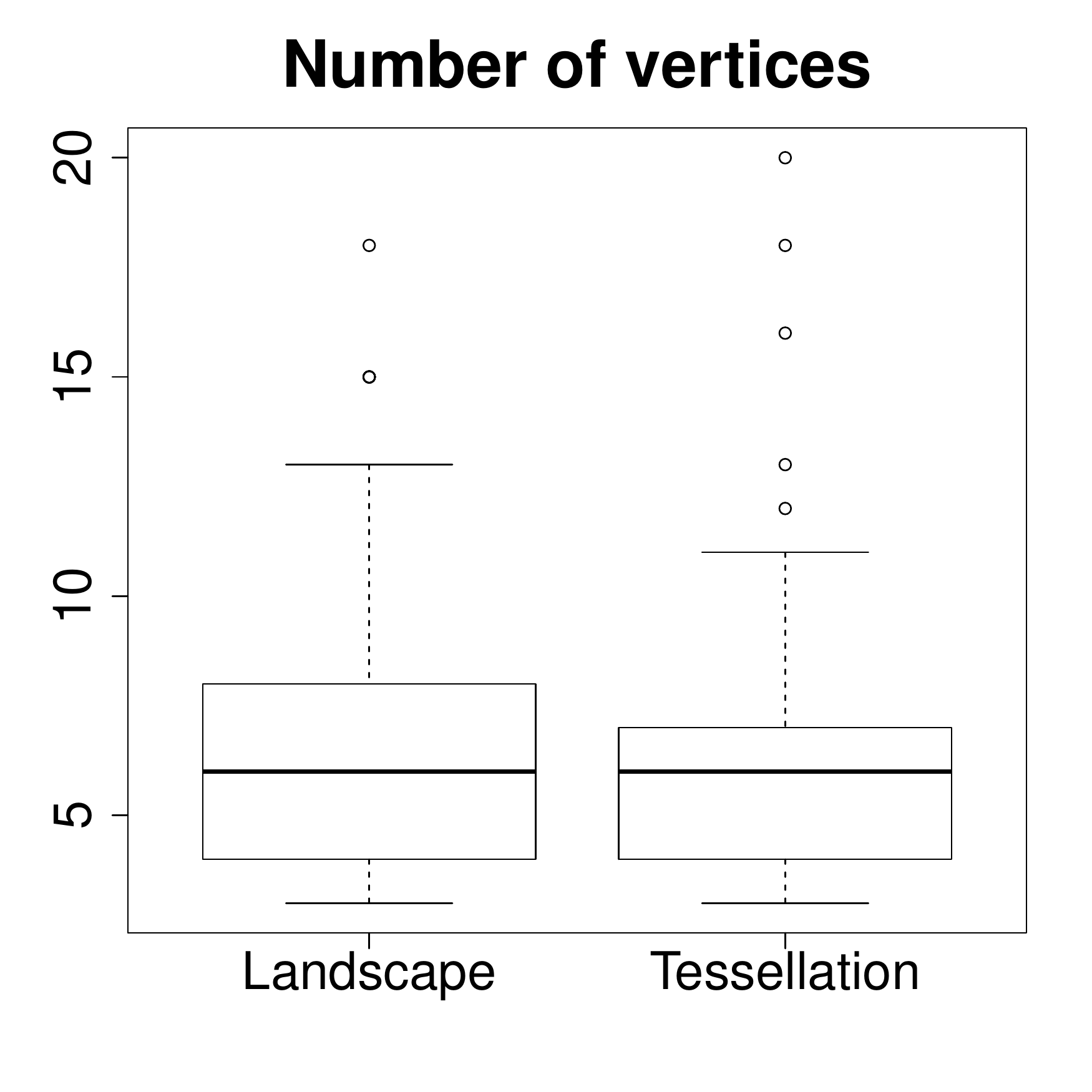}}
\caption{Distribution of the area, the perimeter and the number of vertices of polygons in  the Selommes landscape and
its approximation by a T-tessellation.}
\label{landVstessel}
\end{figure}

Figure (\ref{fig:5}) presents the three landscapes  approximated by T-tessellations. 
The distributions of three basic polygon characteristics in the Selommes landscape and its approximation are compared in Figure
(\ref{landVstessel}). The distributions of perimeter and area are similar before and after
transformation. As for the number of vertices, its median value is comparable but the $75\%$ quantile
and $95\%$ quantile are clearly higher in the original landscape. It is the effect of replacing
closely aligned polygons sides in the original landscape by the representative segments in its approximation.

\section{Gibbsian T-tessellation model}

Let $\Te$ denote a set of all  T-tessellations of a bounded polygonal domain
$W\subset\R^2$. Each tessellation $T\in \Te$ can be defined by the union of the boundaries of its  
cells $\displaystyle E_T$, a closed set with an empty interior. Thus, we can define on  $\Te$ a standard
hitting $\sigma$-algebra for the closed sets,  generated by the events :

\begin{equation*}
\Te_K=\{T\in\Te : \displaystyle E_T \cap K \neq \emptyset\}
\end{equation*}

for $K\in \mathbb{K}$,  a set of all compact subsets of $W$ (see \cite{ref/21}). The corresponding
measurable space will be denoted by $\{\Te,\sigma(\Te)\}$. It should be noted that we can assign to
each tessellation $T\in \Te$ a set of the lines supporting the segments of $T$. Conversely, we can
assign to each line pattern $L$ intersecting $W$ the set of all T-tessellations supported by the lines of $L$. 
This observation leads us to a definition of a counting measure on $\sigma(\Te)$, corresponding to a line pattern $L$.

\begin{definition}
\label{def:2}
Let $L$ be a line pattern intersecting a domain $W$ and let $\Te (L)$ be a set of all T-tessellations supported by the lines of $L$. 
A measure $\mu_L$ on $\sigma(\Te)$ is defined  as:

\begin{equation}
\label{eq:2}
\mu_L(A)=\sum_{T\in\Te (L)}\mathbb{I}_A(T) \text{ }\quad \forall A\in \sigma (T)
\end{equation}
\end{definition}

where $\mathbb{I}_A$ stands for an indicator function of a set $A$. The measure $\mu_L(A)$ counts the tessellations in $A$  supported by the lines of $L$. When $L$ is a realization of a Poisson line process,  the following extension of the previous definition holds: 

\begin{definition}
\label{def:3}

Let $\LL$ be a Poisson line process with unit intensity, restricted to $W$. 
A measure $\mu$ on $\sigma(T)$ is defined as follows:
\begin{equation}
\label{eq:3}
\mu(A)=\frac{1}{z}\E \mu_{\LL}(A)  \text{ }\quad \forall A\in \sigma (T)
\end{equation}
where $z=\E  |\Te (\LL) | $ is a  normalizing constant.
\end{definition}

The model defined by the probability distribution (\ref{eq:3}) was proposed in
\cite{ref/10} and referred to as the Completely Random T-Tessellation Model (CRTT). 
The finiteness of measures (\ref{eq:2}) and (\ref{eq:3}) was proven in \cite{ref/9} and \cite{ref/10}.

The CRTT model can be considered as a reference measure on $\{\Te,\sigma(\Te)\}$. 
Let $s(T)$ be a $d$-dimensional vector of tessellation statistics and let $\theta$ be 
a corresponding parameter vector belonging to a parametric space $\Theta\subset\R ^d$. Let us assume that
each component of $s$ can be written as the sum of local contributions of tessellation vertices, edges, cells
or segments:
\begin{equation}
s_i(T)=\sum_{e\in{\cal{E}}(T)}f_i(e) 
\end{equation}
where ${\cal{E}}(T)$ denotes the set of tessellation elements and $i$ varies from $1$ to $d$. The model that 
favors the tessellations with high (resp. low) values of statistics $s_i(T)$ is given by
the following density function w.r.t. the measure $\mu$:

\begin{equation}
\label{eq:4}
p_{\theta}(T)=\frac{1}{c_\theta}\exp \left (-\sum_{j=1}^d \theta_j s_j(T)\right )
\end{equation}
where $c_\theta$ is an unknown normalizing constant. The scalar product defining the probability density  \ref{eq:4} is denoted by :
\begin{equation}
\label{eq:energy}
U_\theta(T)=\langle\theta,s(T)\rangle
\end{equation}

Model (\ref{eq:4}) is referred as a Gibbs model for T-tessellations and the function $U_\theta(T)$ is the energy
function of the model.  Depending on the chosen statistics, the model makes it possible to control  various T-tessellation features: 
cell area variability, angles between edges, etc.

\begin{wrapfigure}{r}{0.3\textwidth}
  \vspace{-20pt}
  \begin{center}
    \includegraphics[width=0.3\textwidth]{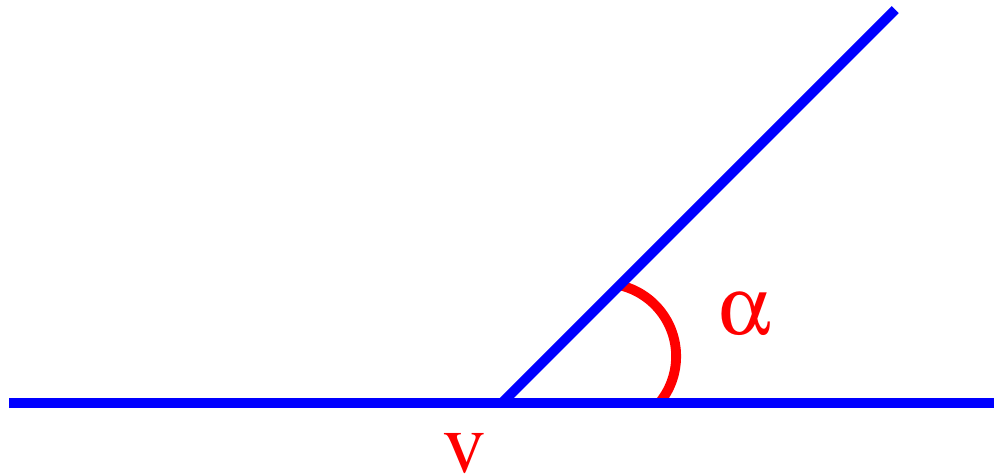}
  \end{center}
  \vspace{-20pt}

\caption{Acute angle between the edges incident to a vertex $v$.}
\label{aa}
 \vspace{-20pt}
\end{wrapfigure}

\begin{example}Let us consider  a Gibbs model with the following statistic : 
\label{ex:1}
\begin{equation}
\label{eq:5}
s_{\sphericalangle}(T)= \sum_{c\in C(T)}\left(\sum_{v\in V(c)}\left(\frac{\pi}{2}-\alpha(v)\right)\right)
\end{equation}

The first sum  in (\ref{eq:5}) runs through a tessellation cells, denoted by $C(T)$. For each cell $c$, the set of its vertices $V(c)$ is considered 
in a second sum. The vertex $v$ contributes to the energy function only if the angle $\alpha(v)$ between the sides
incident to $v$ is acute (see Figure \ref{aa}). The contribution of a vertex $v$ is close to $0$ if the sides are almost perpendicular.
Thus the statistic $s_{\sphericalangle}(T)$ measures a deviation of tessellation cells from rectangles.
\end{example}

A  Metropolis-Hastings-Green algorithm for the simulation of the Gibbs model (\ref{eq:4}) was proposed and its convergence properties
were discussed in  (\cite{ref/10}). It is based on three local updates of a T-tessellation: a split of tessellation cell, a merge of two  cells and
a removal of an edge at the end of a blocking segment, followed by an insertion of an edge  incident 
to a new endpoint of the shortened segment. The algorithm was implemented in the C++ library (\cite{ref/1}).

\begin{figure}
\centering
\subfloat[]{\includegraphics[width=.35\textwidth]{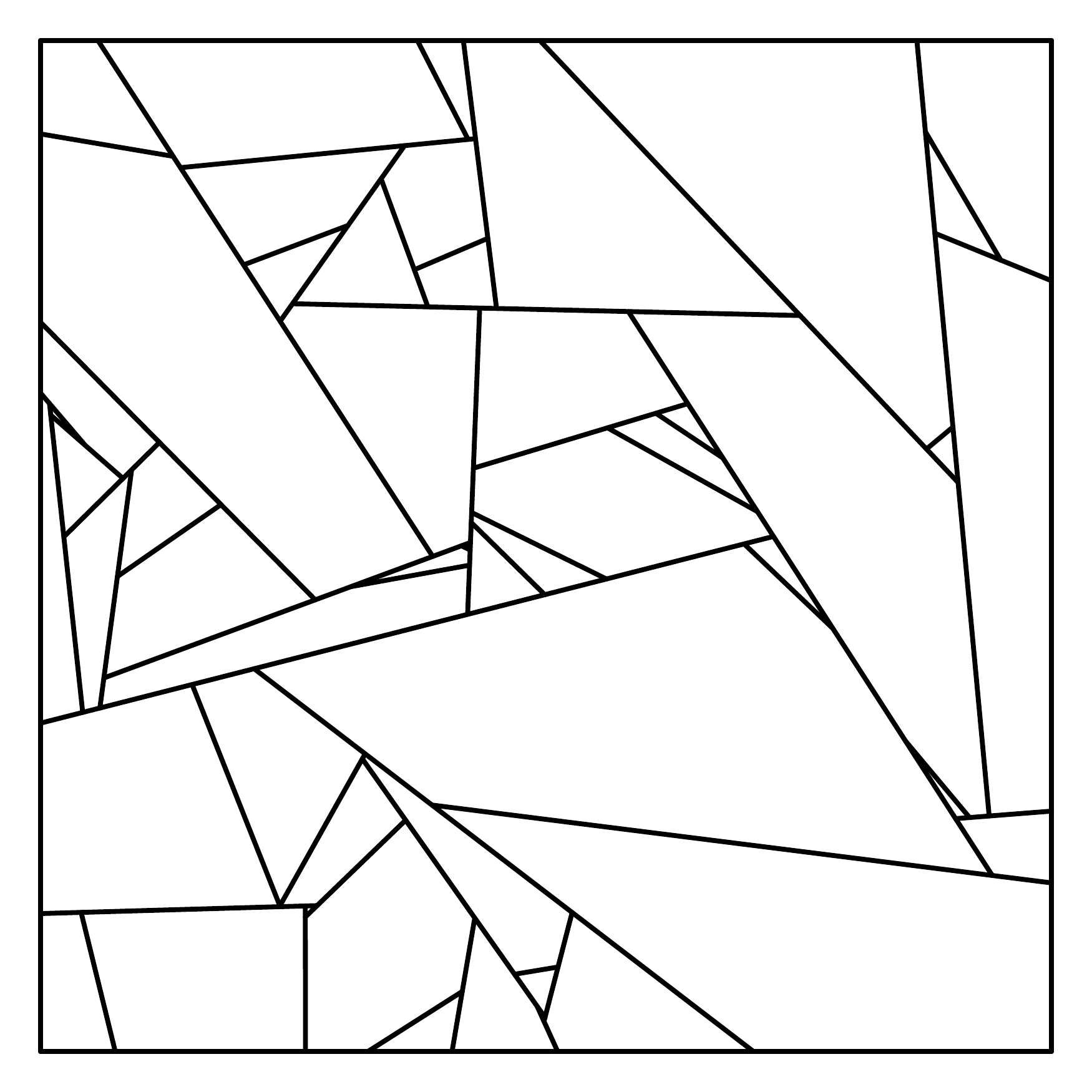}}\quad
\subfloat[]{\includegraphics[width=.35\textwidth]{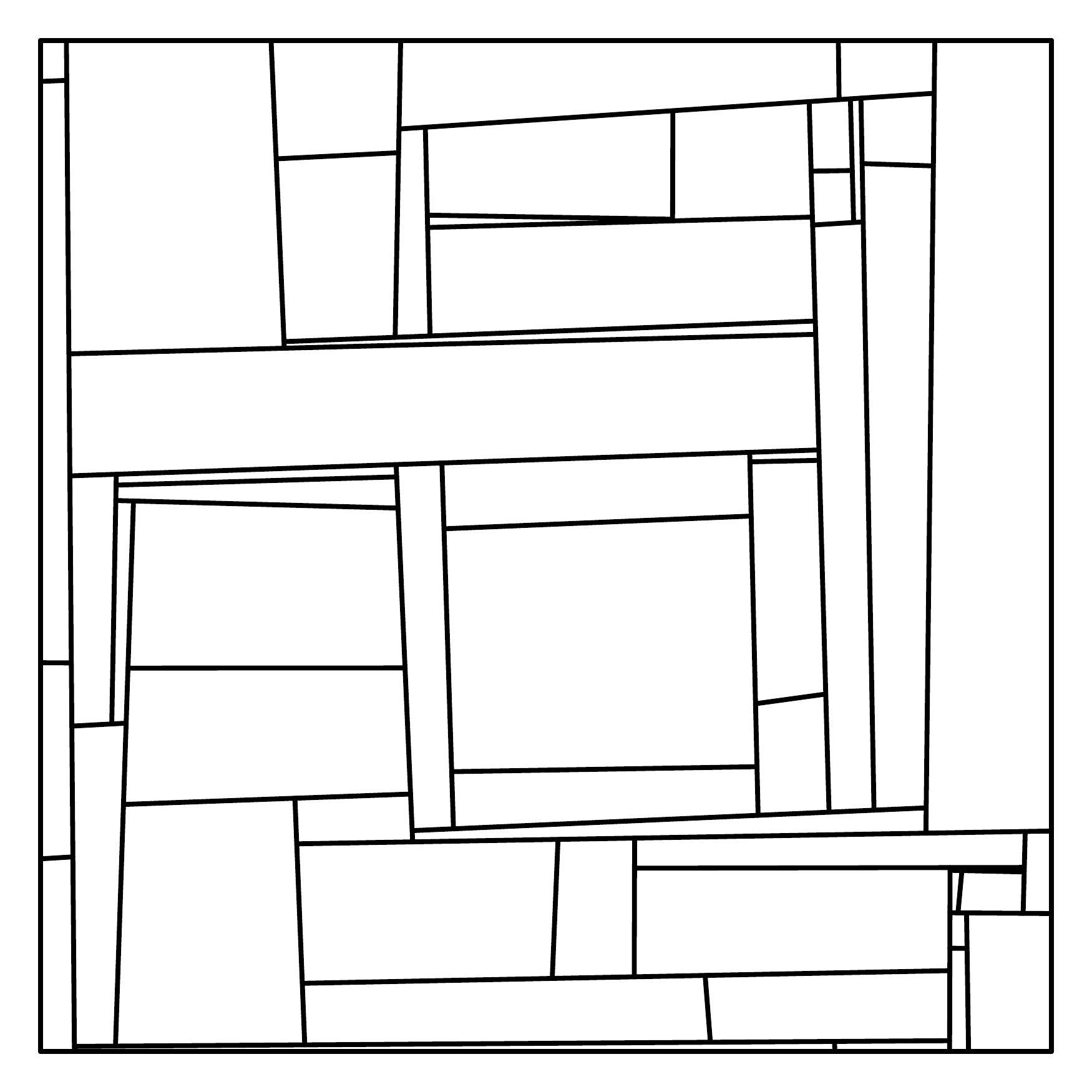}}
\caption{Simulation of: (a) CRTT model (\ref{eq:3}); (b) Gibbs model with angle statistic (\ref{eq:5}).}
\label{fig:1}
\end{figure}

Figure (\ref{fig:1}) illustrates an example of the simulation
of the CRTT model (\ref{def:2}) and of the model with angle statistic $s_{\sphericalangle}(T)$.

\subsection{MCML inference}

Model parameters were estimated by  Monte Carlo Maximum Likelihood method,
proposed in \cite{ref/8} for the families of distributions specified by the unnormalized densities.
The method relies on the possibility of sampling from the targeted distribution.
For the Gibbs model, the log-likelihood ratio against a fixed point $\psi\in \Theta$ is defined as:
\begin{equation}
\label{eq:6}
l(\theta|T,\psi)=\log \frac{p_\theta(T)}{p_{\psi}(T)}=U_\psi(T)-U_\theta(T)-\log\frac{c_\theta}{c_\psi}
\end{equation}

This ratio depends on the ratio of unknown normalizing constants,  which can be expressed as:
\begin{equation}
\label{eq:7}
\frac{c_\theta}{c_\psi}=\E \exp\left (U_\psi(T) -U_\theta(T)\right )
\end{equation}
when the expectation in (\ref{eq:7}) is calculated under the known distribution $P_\psi$. Let ${\bf T}_n=(T_1,\ldots,T_n)$ 
be a sample  simulated from $P_\psi$. The expected value in  (\ref{eq:7}) can be estimated by the empirical mean:
\begin{equation}
\label{eq:8}
\frac{c_\theta}{c_\psi}\simeq\frac{1}{n}\sum_{i=1}^n \exp\left (U_\psi(T_i) -U_\theta(T_i)\right )
\end{equation}

The estimate of the unknown log-likelihood ratio (\ref{eq:6}) is obtained by substituting 
the ratio of normalizing constants with its approximation (\ref{eq:8}). It
is called a Monte Carlo Likelihood (MCL) function and denoted  by $\hat{l}_n(\theta)$. The maximizer  $\hat{\theta}_n$ of the MCL
function approximates the unknown value of the Maximum Likelihood estimate $\hat{\theta}$. The asymptotic properties
of $\hat{l}_n$ and $\hat{\theta}_n$ were derived in \cite{ref/7} and \cite{ref/14}.

The likelihood of exponential family models is convex. Hence the optimization of its Monte Carlo approximation is justified. 
The method holds for every value of $\psi$, but in practice the choice of this parameter has an impact on the quality of the estimation. 
If  $\psi$ is far from the exact estimate, MCL approximation is poor and $\hat{\theta}_n$ will be far from $\hat{\theta}$ as well. One possible solution 
to this problem is to iterate the optimization algorithm. The trust region method applied here for calculating  $\hat{\theta}_n$ (see \cite{ref/28}) starts from an arbitrary parameter $\psi$ and maximizes quadratic approximation of $\hat{l}_n$ over a region $\Delta$ around $\psi$, yielding a new parameter value. The size of the region $\Delta$ is adjusted in order to obtain a correct approximation of $\hat{l}_n$ by a quadratic model.

\begin{figure}
\centering
\subfloat[]{\includegraphics[width=.33\textwidth]{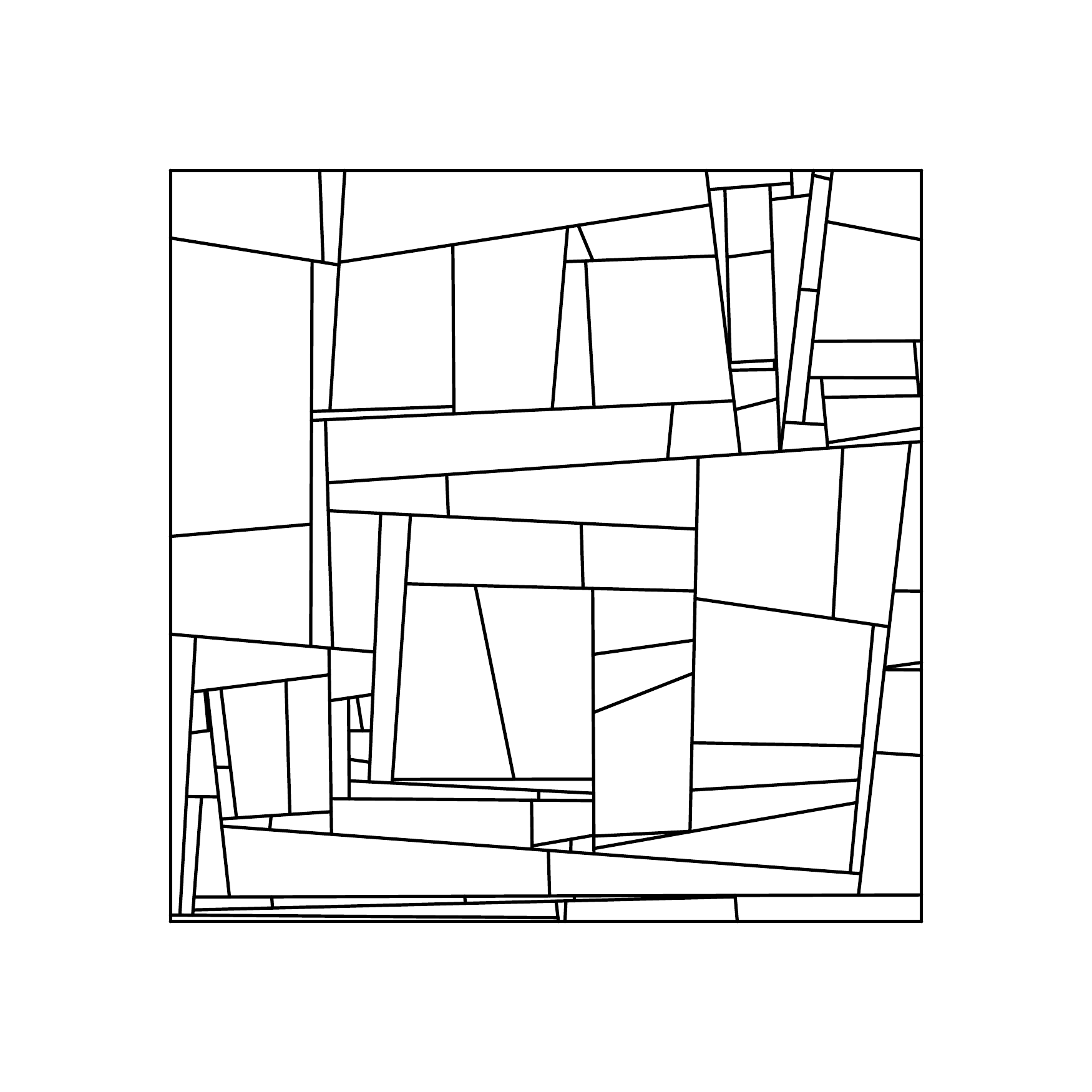}}
\subfloat[]{\includegraphics[width=.33\textwidth]{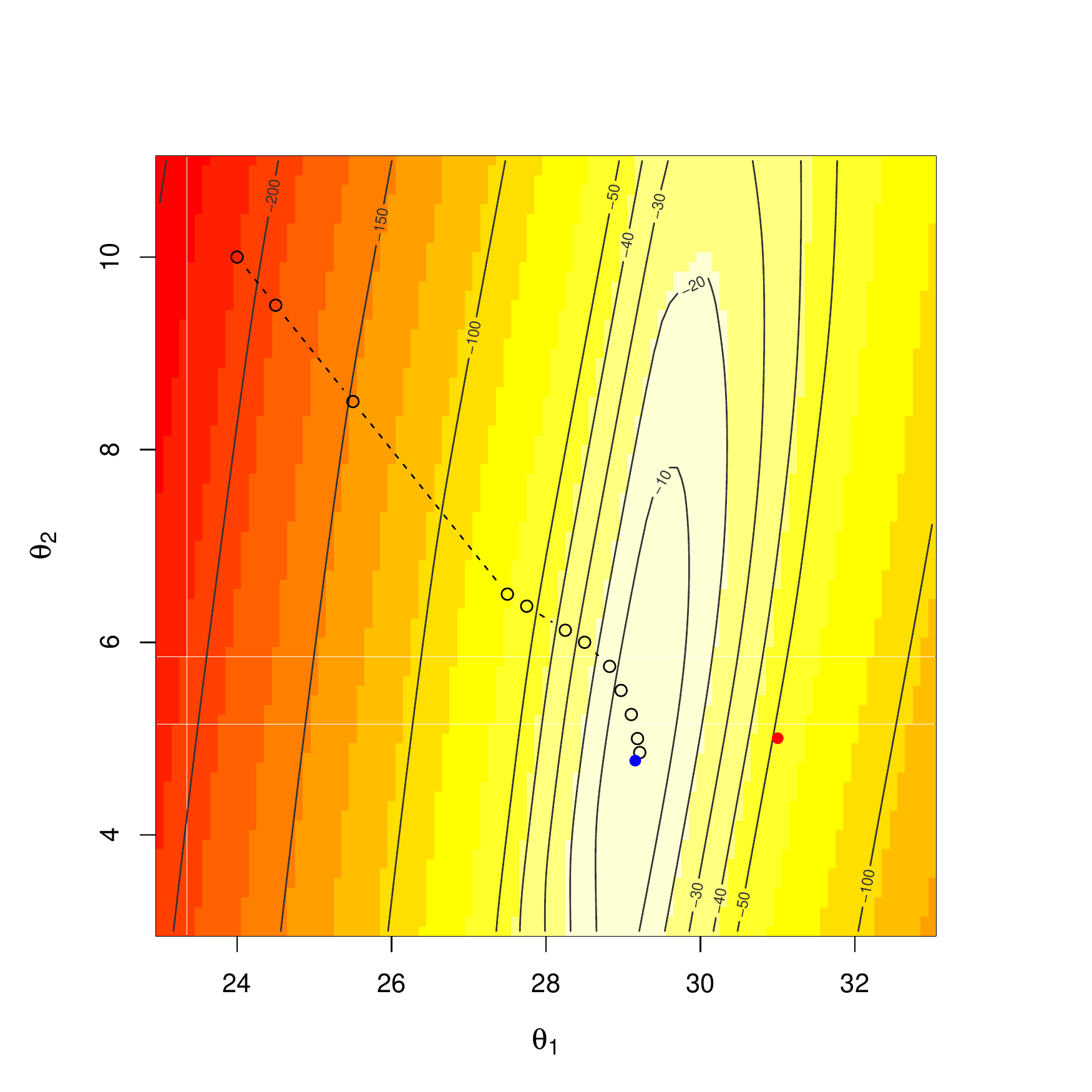}}
\subfloat[]{\includegraphics[width=.33\textwidth]{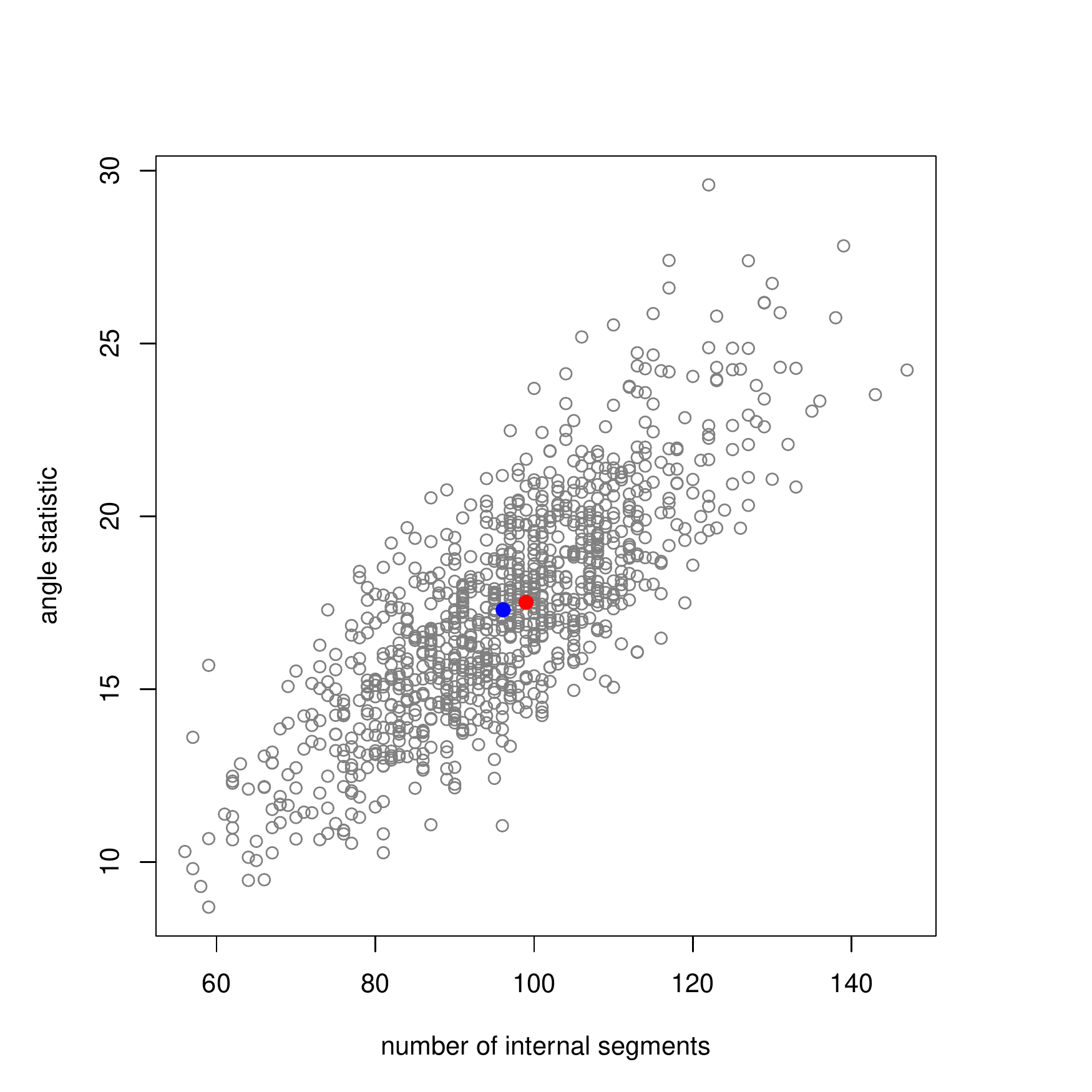}}
\caption{Estimation of parameters of the model (\ref{sim}). (a) Training tessellation $T_0$ simulated with $\theta=(31,5)$. (b) Contour plot of MCL function. The true parameter is represented by a red dot. The algorithm starts at $\psi=(24,10)$ and converges after $12$ iterations at $\hat{\theta}_n=(29.15,4.76)$ (blue dot).  (c) Scatter plot of the values of  energy statistics calculated for $1000$ simulations of the fitted model. The blue and the red dot represent respectively the mean and the observed  values of the statistics. }
\label{fig:MCML}
\end{figure}

Figure \ref{fig:MCML} illustrates the algorithm for calculating the estimates of a two-parameter model with an energy function:
\begin{eqnarray}
\label{sim}
U_\theta(T)=-\theta_1 n_s(T)+\theta_2s_{\sphericalangle}(T) 
\end{eqnarray}
where $n_s(T)$ is the number of internal segments of $T$ and $s_{\sphericalangle}(T)$ is the angle statistic already defined
in  example (\ref{ex:1}). The model was fitted to the tessellation $T_0$ simulated with parameter $\theta=(31,5)$  in the unit square (see Fig. \ref{fig:MCML}a). The algorithm started from $\psi=(24,10)$ and reached the maximum after $12$ iterations. The calculation of a current value of the estimate
was based on a Monte Carlo sample of size $400$.  The distribution of two statistics calculated over $1000$ runs of the fitted model appears to be correctly centered on the observed values (see Fig. \ref{fig:MCML}c).

\begin{figure}
\centering
\subfloat[]{\includegraphics[width=.36\textwidth]{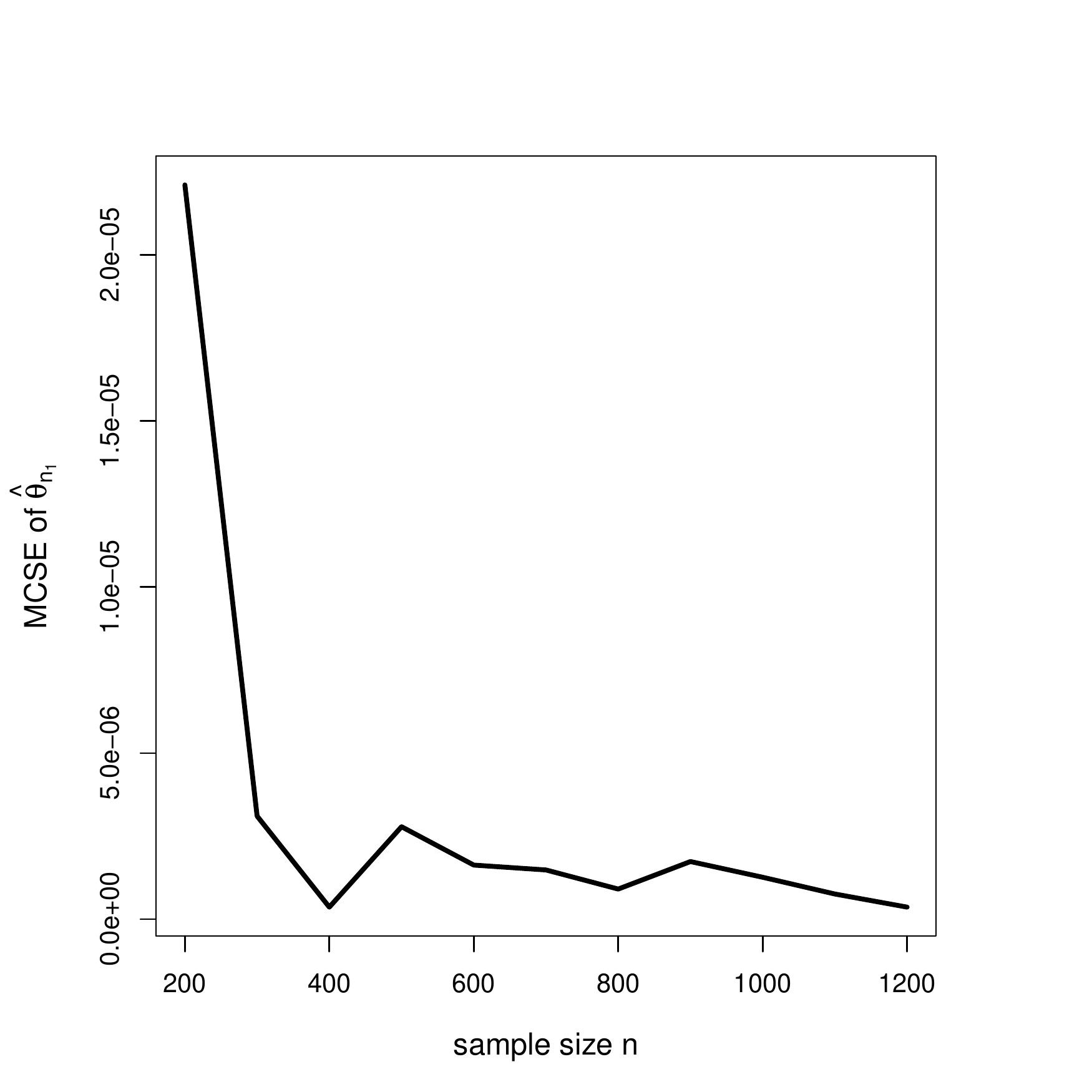}}
\subfloat[]{\includegraphics[width=.36\textwidth]{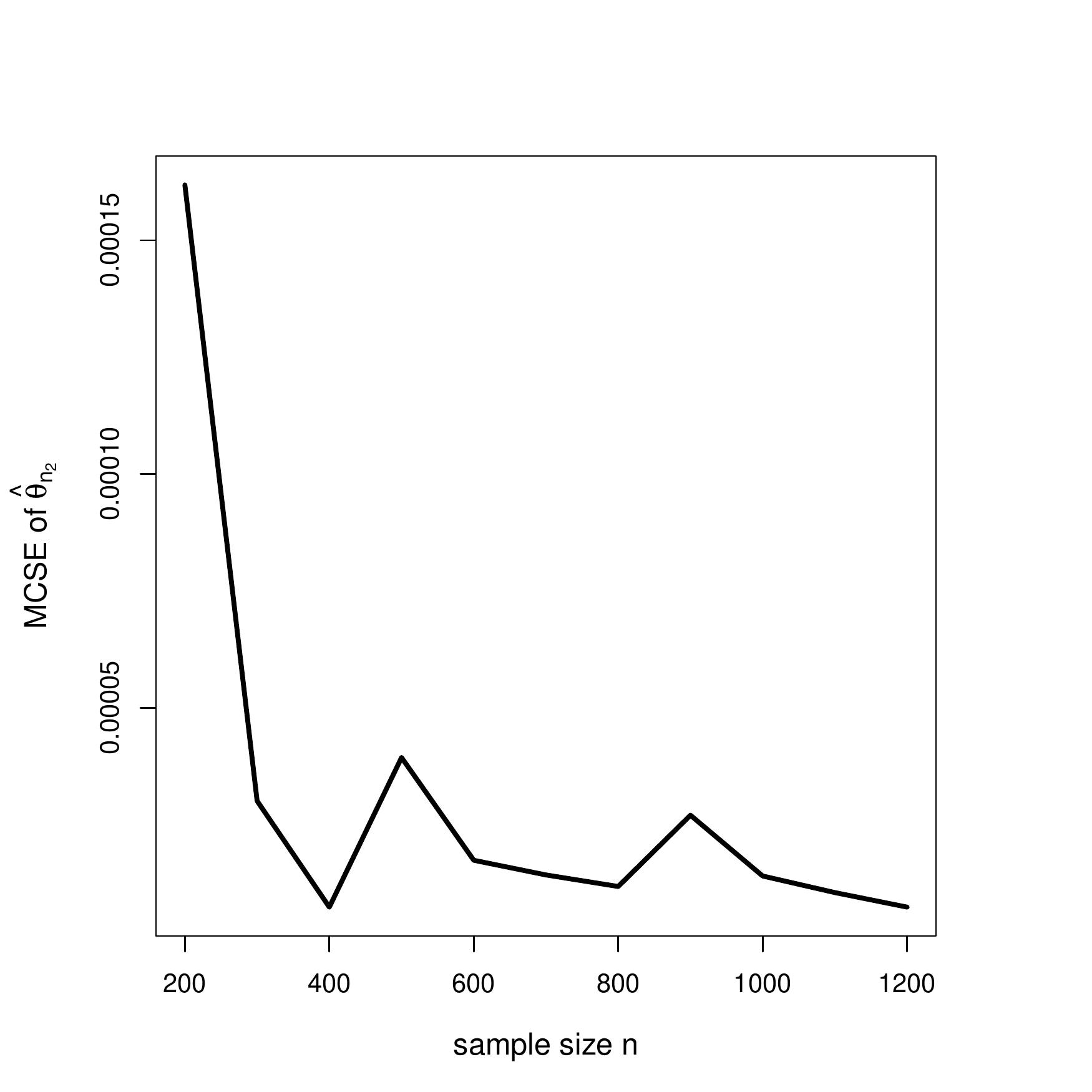}}
\caption{Monte Carlo standard error of $\hat{\theta}_{n_1}$  (left panel)  and $\hat{\theta}_{n_2}$ (right panel) as a function of  sample size $n$.}
\label{fig:4}
\end{figure}

The standard error of the  Maximum Likelihood estimate $\hat{\theta}$ for model 
(\ref{sim}), calculated from the Monte Carlo approximation of the Fisher Information Matrix with $n=400$, was equal to $(0.0135,0.2900)$.
The asymptotic results given by Geyer in (\cite{ref/7}) make it possible to  estimate the standard error of the approximation $\hat{\theta}_n$, referred to as the
Monte Carlo Standard Error (MCSE). Figure (\ref{fig:4}) shows the graphs of MCSE($\hat{\theta}_{n_1}$) and MCSE($\hat{\theta}_{n_2}$) 
as the sample size grows from $n=200$ to $n=1200$.  The highest value of MCSE($\hat{\theta}_n$) was equal
to $(0.000021,0.00016)$ for the sample size $n=200$. The MCSE($\hat{\theta}_n$) decreased rapidly as the sample size varied from $200$ to $600$.
Beyond this threshold no significant improvement of the accuracy of $\hat{\theta}_n$ was observed.

\section{Gibbs model for landscapes comparison and simulation}

The epidemic models of plant diseases are studied at the landscape scale. Landscape features are likely to play
a role in the epidemic spread, interacting with  other variables determining the dynamics of the host-pathogen
system. The impact of variables is evaluated through numerical experiments (see \cite{ref/19}). 
Landscape effect is measured by comparing the model's outputs calculated for the landscapes
contrasted with respect to the targeted features, like in factorial experiments. The design of such experiments
requires a set of landscapes that differ with respect to the variables of interest. If the access to the landscape databases
is somehow restricted, simulations of simplified landscape representations can be an alternative approach.

The Gibbs model can be a useful tool for yielding contrasted tessellation patterns that inherit the observed landscapes
features. Let us illustrate this approach on the example of three landscapes in (\ref{fig:5}). The first step is to 
find out what the characteristics are that differ between the landscapes, fitting a guess model to landscape
approximations and comparing the estimates of model parameters.

\begin{wrapfigure}{r}{0.2\textwidth}
   \vspace{-30pt}
  \begin{center}
    \includegraphics[height=3cm]{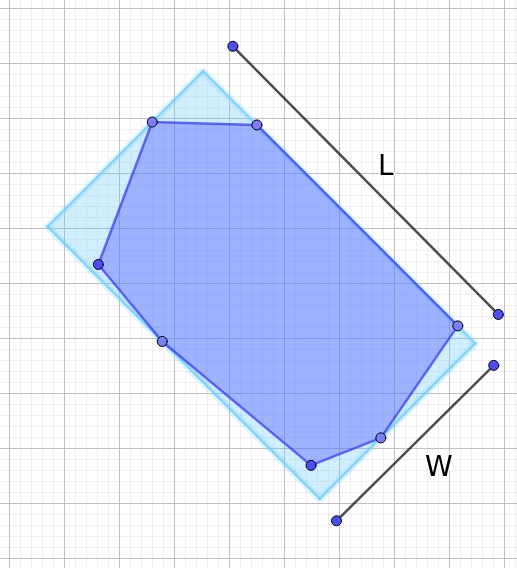}
  \end{center}
\caption{Length-to-width ratio of minimal bounding rectangle.}
\label{l2w}
\end{wrapfigure}

The guideline for choosing the model statistics is therefore their ability to capture the contrasts between the observed patterns.
Based on these considerations, the candidate model with the following statistics was proposed first:

\begin{itemize}
\item the number of tessellation cells $s_1(T)$, accounting for the differences in tessellation scales;
\item the sum of squared cell areas $s_2(T)$, measuring
      the departure from the patterns with cells with equal areas, minimizing the value of the statistic;
\item the angle statistics $s_3(T)$, measuring the departure from orthogonal patterns. This statistic, already introduced in  
      example (\ref{eq:5}), was renamed here so that all of the statistics would have a common notation;

\item the number of long cells $s_4(T)$. A cell is said to be long if the length-to-width ratio of its smallest enclosing rectangle (see Fig. \ref{l2w})
      is greater than a threshold $l_0$. In this case the value of $l_0$ was fixed at $4$, the value close to the median length-to-width ratio in the BVD landscape.
\end{itemize}

Let $T_1$, $T_2$ and $T_3$ be the observed tessellations, approximating the landscapes represented in Figure ($\ref{fig:5}$). Let a tessellation $T_i$ follow a Gibbsian model given by the energy: 

\begin{equation}
\label{LeModel}
U_{\theta_i}(T_i)=\sum_{j=1}^4\theta_{i,j}s_{j}{(T_i)}
\end{equation}

for $i \in \{1,2,3\}$. Table (\ref{estimates}) gives Monte Carlo Maximum Likelihood estimates of model parameters  together with their $95\%$ confidence sets,
based on the approximation of the Fisher Information Matrix inverse. The size of the Monte Carlo sample was fixed at $n=500$.
\begin{table}[b]
\label{estimates}
\begin{center}      
\begin{tabular}{|l|c|c|c|c|}
\hline
Landscape $T_i$& $\hat{\theta}_{i,1} $ & $\hat{\theta}_{i,2}$& $\hat{\theta}_{i,3}$ & $\hat{\theta}_{i,4}$ \\
\hline
\hline
Selommes & -1.98  & 182  & 2.22  & 0.27 \\
         &  (-2.23,-1.73) & (55,309) & (1.92,2.50) & (-0.05,0.61) \\
&&&&\\
Kervidy  & -1.69  & 80    & 1.26  & 0.05 \\
         &   (-1.90,-1.49) & (-25,187) & (1.04, 1.48) & (-0.28,0.38) \\
&&&&\\
BVD      & -2.08  & 2518 & 2.18  & -0.58 \\
         & (-2.32,-1.83) & (1217,3819) & (1.89,2.47) & (-0.83,-0.48)\\
\hline
\end{tabular}

\end{center}

\caption{MCML estimates of the parameters of  model  (\ref{LeModel}) and their $95\%$-confidence sets.}

\label{estimates}
\end{table}

The model fitted to the BVD landscape tends to generate the tessellations close to the patterns with equally sized cells (high positive value of $\hat{\theta}_{2}$)
and orthogonal segments (positive value of $\hat{\theta}_{3}$).The model favors the patterns with a  number of long cells greater than in other landscapes
and greater than in the reference model (negative value of $\hat{\theta}_{4}$). The model fitted to the Kervidy landscape is closest to completely random T-tessellation. 
The estimates of parameters ${\theta}_{2}$ and ${\theta}_{4}$ are non-significantly different from zero. The penalty for the acute angles between the segments ($\hat{\theta}_{3}$) is positive, but its value is smaller than the estimates calculated for the Selommes and BVD landscapes. The model fitted to Selommes data generates the most rectangular patterns with the smallest number of long cells.

\begin{figure}
\centering
\subfloat[]{\includegraphics[width=.31\textwidth]{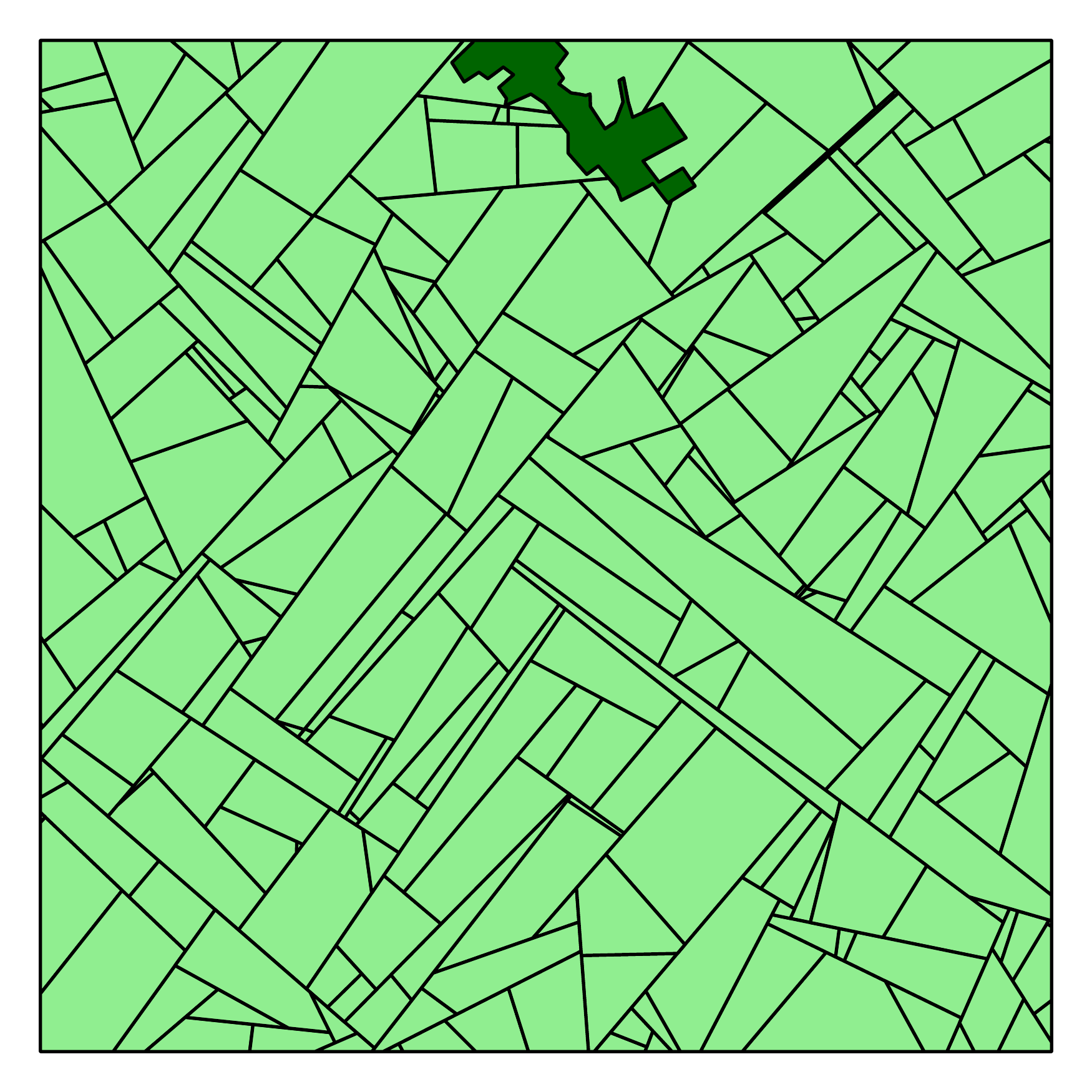}}\quad
\subfloat[]{\includegraphics[width=.31\textwidth]{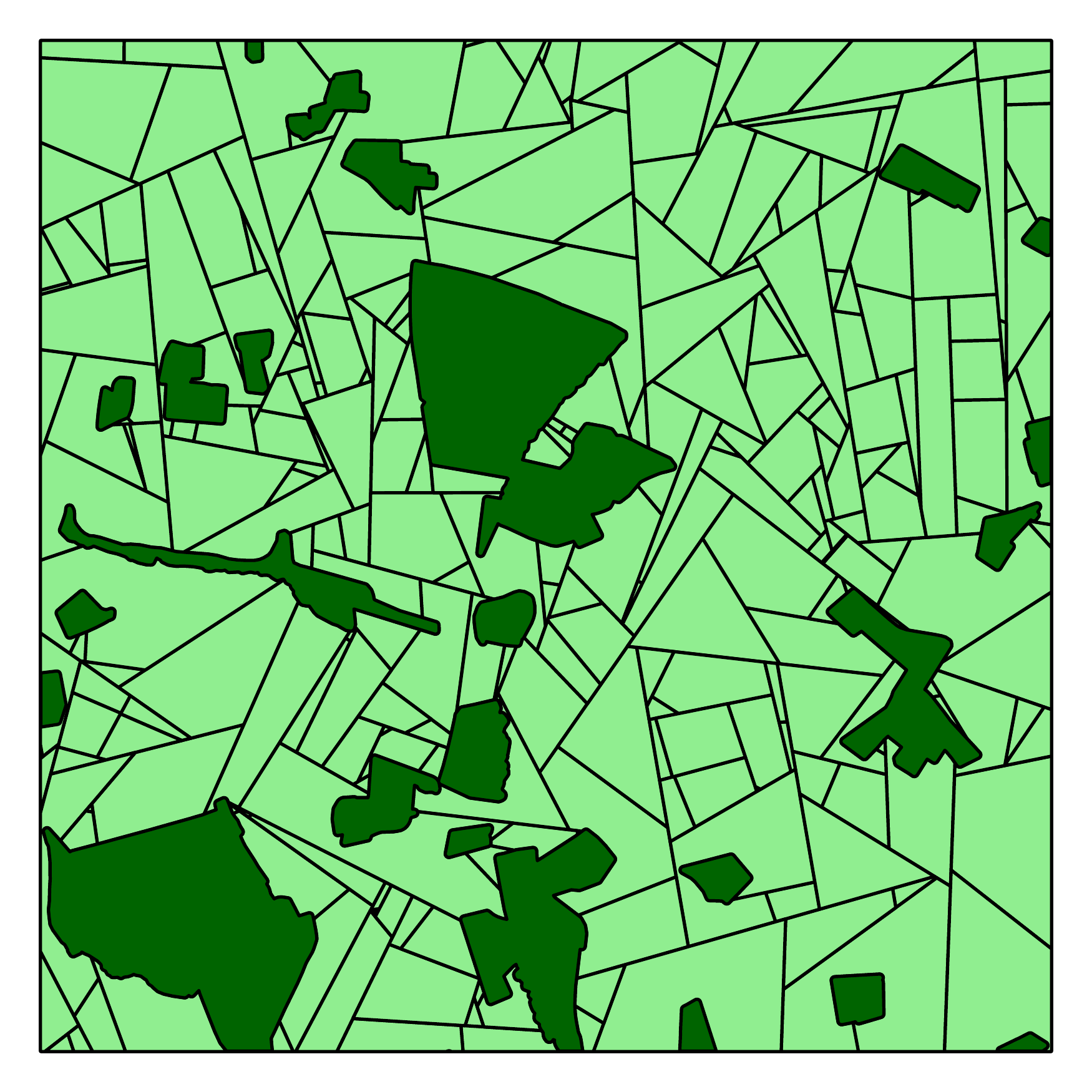}}\quad
\subfloat[]{\includegraphics[width=.31\textwidth]{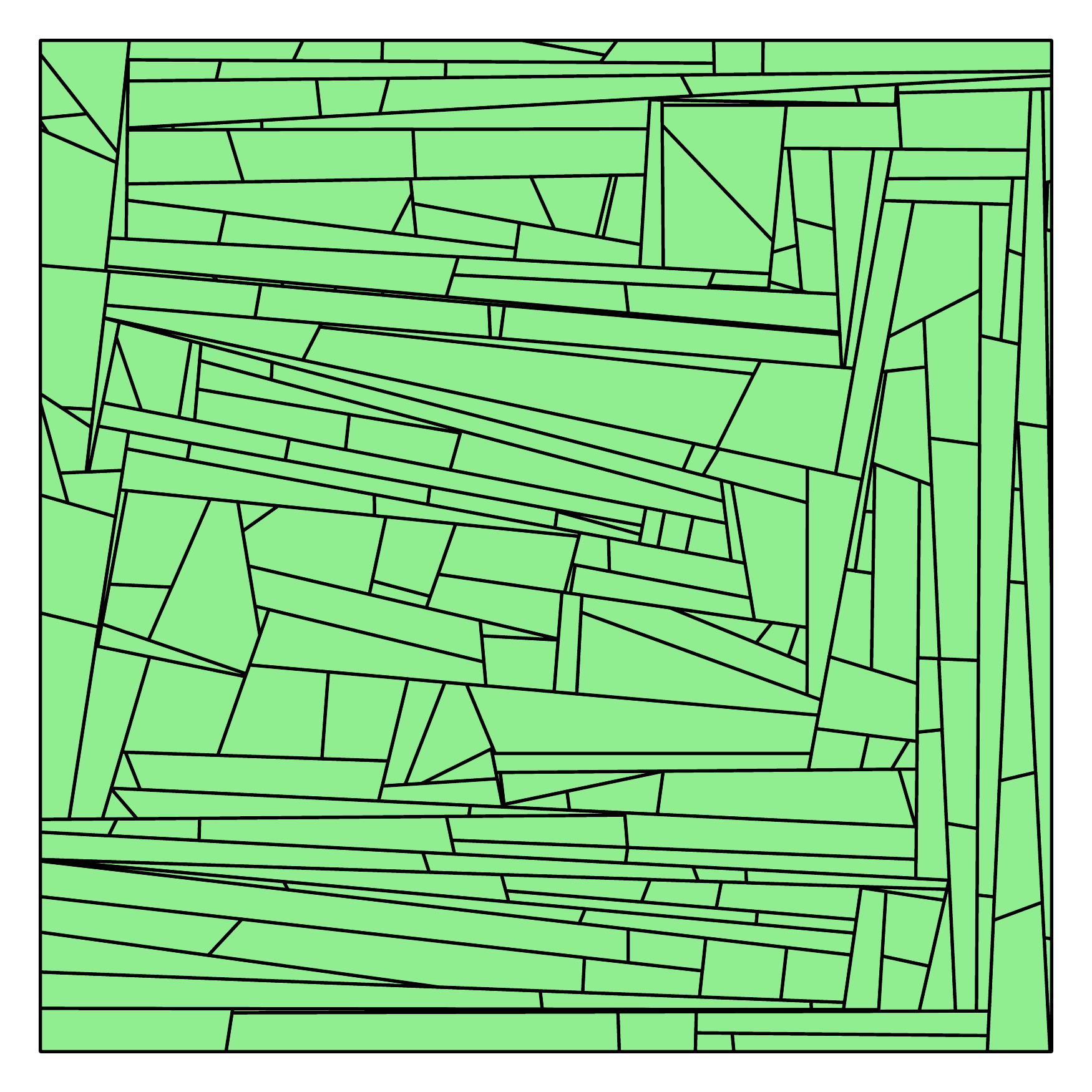}}\quad
\caption{Simulations of Gibbs model (\ref{LeModel}) fitted to (a) Selommes, (b) Kervidy and (c) BVD landscapes.}
\label{fig:simul}
\end{figure}

Figure (\ref{fig:simul}) gives the examples of model simulations for the three landscapes. According to the model properties, the mean values of
the statistics calculated over a sample of tessellations simulated from (\ref{LeModel}) are centered on the values observed in the landscapes.
In this sense, the model is able to generate simplified landscape representations that inherit selected landscape features.

\subsection{Empty-space function for testing goodness-of-fit}

The empty-space function $F$ is a summary statistic commonly applied in spatial point pattern analysis to characterize
the observed patterns and to check point process model validity  (see  \cite{ref/2}). The function definition can be extended
to the other random sets, e.g. line segments and linear networks (see \cite{ref/27}, \cite{ref/5}). For a random stationary tessellation ${ T}$, the empty-space
function is defined as the cumulative distribution function of the shortest distance $d(u,T)$ from an arbitrary point $u$ in the tessellation domain to $T$:

\begin{equation}
\label{eq:F}
F_{ T}(r)=P(d(u,T)\leq r)
\end{equation}

Since  $T$ is stationary, the definition (\ref{eq:F}) does
not depend on a particular choice of  $u$. The function $F_{ T}(r)$ can be interpreted as the average fraction of the area of the region obtained
by the dilation of tessellation $T$ by a disc with a radius $r$. Consequently, for tessellations with cells that have regular shapes and sizes, such as grid patterns,  $F_{ T}$ will tend to grow more rapidly than, for example, in the CRTT model. The empty-space function is thus a useful tool for comparing the observed patterns 
between themselves or with the patterns generated by a candidate model. The border-corrected estimate of $F_{T}(r)$ is calculated over the points
sampled in the tessellation domain and separated from the border by a distance greater than or equal to $r$: 
\begin{equation}
\label{eq:defF}
\hat{F}_{ T}(r)=\frac{\#\{u_i: d(u_i,T)\leq r \cap d(u_i,\partial W)\geq r \}}{\#\{u_i:  d(u_i,\partial W)\geq r\}}
\end{equation}
where $\partial W$ denotes the border of the tessellation domain $W$ (see \cite{ref/5}).

The global envelope test for model checking compares a functional statistic of the observed  pattern with its counterparts
calculated for the simulations of the null model (see \cite{ref/16}). The test is applied for checking point process models, but its extension to
random tessellation models is straightforward. The principle of the test procedure is to reduce a tessellation pattern $T$ to a random variable $X=X(T)$ and to
compare the distribution of $X$ for the observed pattern with its distribution simulated from a null model. Let $X_{\text{obs}}$ denote the value of $X$ for 
the observed tessellation $T_{\text{obs}}$ and let $X_1,\ldots,X_{m-1}$ be the values of $X$ calculated for a sample of tessellations $T_1,\ldots, T_{m-1}$ generated by a null model. The tessellations $T_1,\ldots, T_{m-1}$ are assumed to be independent of each other and independent of the observed pattern.
Let $P_x$ denote the  distribution of the i.i.d. sample $X_1,\ldots,X_{m-1}$. A null hypothesis states that:
\begin{equation}
\label{H0}
H_0 :  X_{\text{obs}}\sim P_x
\end{equation}

Let $R_{\text{obs}}$ be the rank of $X_{\text{obs}}$ in a pulled sample $X_{\text{obs}},X_1,\ldots,X_{m-1}$. Under the null hypothesis
this is an i.i.d sample and the following equality holds:

\begin{equation*}
P_{H_0}(R_{\text{obs}} > m-1)=\frac{1}{m}
\end{equation*}
Thus, if $X_{\text{obs}}$ is greater than a maximal value of $X_i$ than the null hypothesis (\ref{H0}) is rejected at a significance 
level $\alpha=\frac{1}{m}$.

The choice of the variable $X(T)$ is subject to the constraint that  its large positive values  are unlikely under
a null hypothesis.  Different variants of envelope test were proposed depending on the definition of $X(T)$ (see \cite{ref/2}). In the Maximum Absolute Deviation test  based on the empty-space function, the variable $X(T)$ is defined as the maximal absolute difference between the estimate of $F_T$  and its expected value under the null hypothesis:
\begin{equation}
\label{MAD}
X(T)=\max_{0\leq r \leq R}|\hat{F}_T(r)-{F}_{\text{ref}}(r)|
\end{equation}
The maximum in (\ref{MAD}) is taken over a range of distances from $0$ to $R$ units. If ${F}_{\text{ref}}$ in  formula (\ref{MAD}) cannot 
be calculated explicitly, it can be estimated from the simulations of a null model.
Testing the null hypothesis (\ref{H0}) with the MAD test is equivalent to checking if $\hat{F}$ calculated for the observed tessellation  falls
inside the interval:
\begin{equation}
\label{env}
\left({F}_{\text{ref}}(r)-\max_iX_i,{F}_{\text{ref}}(r)+\max_iX_i\right) 
\end{equation}
for every $r$ in $(0,R)$. If this is the case, there is no evidence for rejecting $H_0$.

\begin{figure}
\centering
\subfloat[]{\includegraphics[width=.33\textwidth]{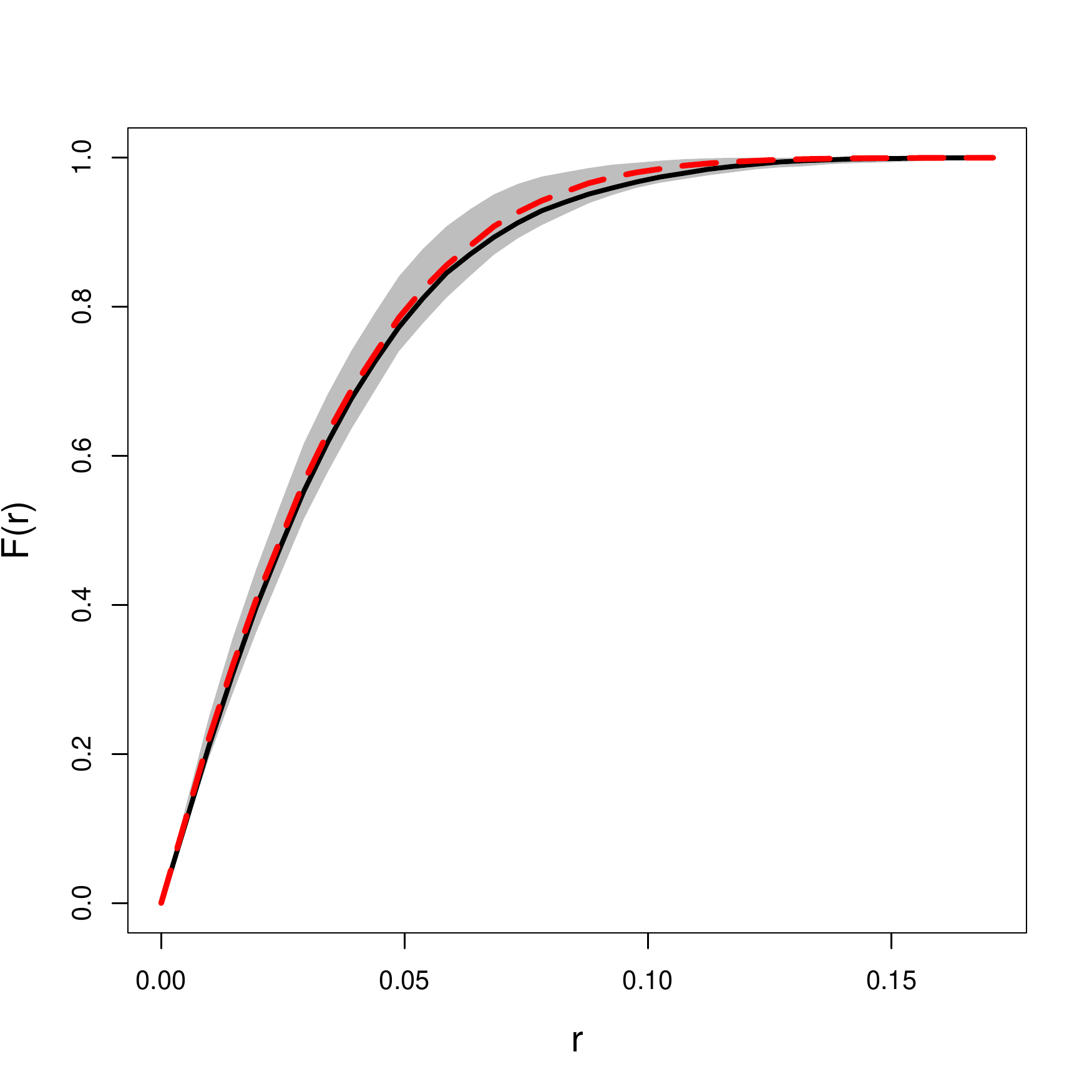}}
\subfloat[]{\includegraphics[width=.33\textwidth]{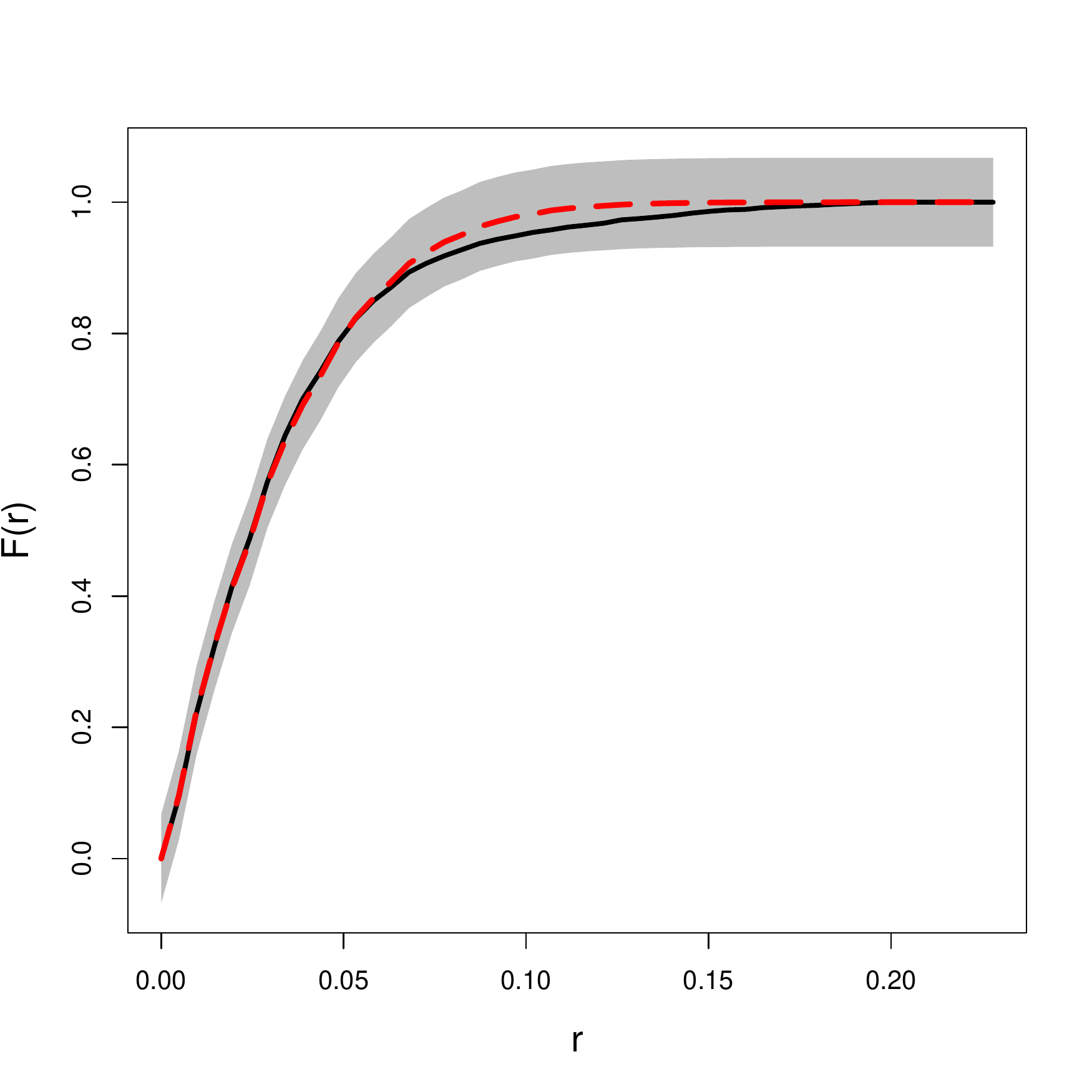}}
\subfloat[]{\includegraphics[width=.33\textwidth]{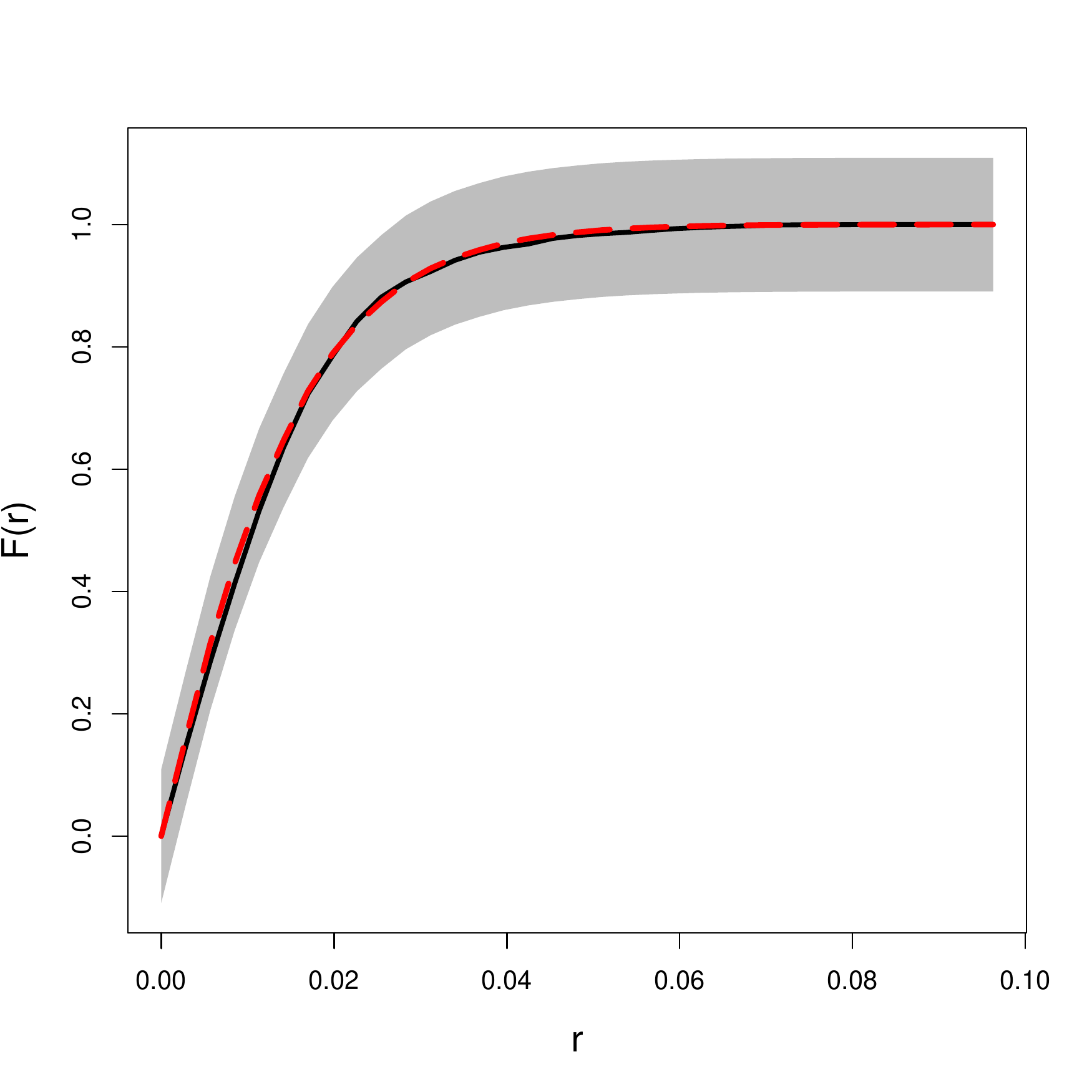}}

\caption{Simultaneous envelopes (\ref{env}) for the empty-space function: (a) Selommes landscape; (b) Kervidy landscape; (c) BVD  landscape. Estimates of the 
empty-space function for the observed patterns are plotted in black. The dashed red curves represent the estimates of $F_{\text{ref}}$ for   model (\ref{LeModel}).}
\label{fig:emptyspace}
\end{figure}

The test was applied to check the goodness-of-fit of model (\ref{LeModel}). Figure {\ref{fig:emptyspace}} shows the envelopes (\ref{env})
based on $m-1=499$ model simulations. The estimated $F$-function for the observed patterns falls
entirely inside the envelopes for each landscape. At the significance level $\alpha=\frac{1}{500}$ there is no evidence against the null hypothesis that landscape patterns follow the Gibbs model (\ref{LeModel}) with the parameters values given by Table (\ref{estimates}). The corresponding p-values of the test are:
$0.53$ for the Selommes, $0.38$ for the BVD and $0.20$ for the Kervidy landscape.

\section{Conclusions}

Random T-tessellations following the Gibbs model proposed in (\cite{ref/10}) may represent
a broad scope of spatial patterns. This paper completes the model with a statistical
inference tested on the example of agricultural landscapes, approximated by T-tessellations.
The proposed landscape model highlights the differences between the observed patterns.
Model parameters are estimated by the Monte Carlo Maximum Likelihood method. Confidence sets
calculated from MC approximation of the Fisher Information Matrix enable parameters
comparison. The goodness-of-fit of the model is assessed by the global envelope test for the
empty-space function.

The framework of Monte Carlo Likelihood inference may  give rise to the  development of 
variable selection methods for the Gibbs model. Starting from a set of candidate variables, 
a stepwise procedure can be  considered for variables inclusion and elimination, based on Monte Carlo approximation 
of Likelihood tests. From a practical view point it should be accompanied by the development
of more efficient computational methods for finding $\hat{\theta}_n$. Indeed, the calculation of
the estimate becomes cumbersome as the number of variables grows. 

Strong correlations between the model statistics may be at the origin  of the difficulties in model
fit. The example of the model fitted to the BVD landscape illustrates  this problem.
The model accounts for a high number of long cells and, at the same time, favors the patterns 
with equally distributed cell areas. In order to satisfy both criteria the model  tends
to generate  the cells with extremely elongated shapes, as illustrated in Figure (\ref{fig:simul}).
Consequently, the model alters the mean elongation of tessellation cells, making it difficult to 
fit the model while controlling the values of this statistic at the same time. The question of
how to select a set of tessellation statistics to avoid correlation issues is of major interest 
for statistical inference.

Finally, the generalization of the model to polygonal tessellations with no restrictions on the vertices type
would be of great practical interest as well. The source of bias due to the approximation of the original pattern
by a T-tessellation could be avoided in this way.

\section*{Acknowledgments}
We are grateful to Claire Lavigne and Sylvain Poggi for access to the landscape datasets and
for their precious advice.

This work was supported partly by the French PIA project “Lorraine Université d’Excellence” (ANR-15-IDEX-04-LUE)
and by the Metaprogram  SMACH  (Sustainable Management   of   Crop   Health, http://www.smach.inra.fr/)   of  
the French National Research Institute for Agriculture, Food and Environment (INRAE).

\bibliographystyle{plain}
\bibliography{GibbsForLandscape}

\end{document}